\documentclass[twocolumn,aps,prx,footinbib,floatfix,
longbibliography,superscriptaddress,10pt]{revtex4-2}

\usepackage{amsmath}
\usepackage{amssymb}
\usepackage{bm}
\usepackage{mathtools}
\usepackage{xcolor}
\usepackage{graphicx}
\usepackage[normalem]{ulem}
\definecolor{webblue}{rgb}{0,0,0.6}
\usepackage[unicode=true,colorlinks=true,linkcolor=webblue,citecolor=webblue,urlcolor=webblue]{hyperref}
\usepackage[utf8]{inputenc}
\usepackage[T1]{fontenc}
\usepackage{pdfpages}
\usepackage{pgffor}

\usepackage[capitalize]{cleveref}
\usepackage{makerobust}
\usepackage{braket}
\usepackage{bbold}
\usepackage{comment}
\usepackage{soul}

\newcommand{\bvec}[1]{\boldsymbol{#1}}
\newcommand{\tWSe}{tWSe\textsubscript2}
\newcommand{\vdagger}{{\vphantom{\dagger}}}
\newcommand{\matrixOne}{\mathbb{1}}

\newcommand{\makeauthor}[2]{\newcommand{#1}[1]{{%
  \protect%
  \color{#2}{%
    \bfseries\begingroup\escapechar=-1\edef\x{\endgroup\string#1}\x:%
  } ##1}}%
  \MakeRobustCommand#1}
\newcommand{\makeauthorNew}[2]{\newcommand{#1}[1]{{%
  \protect%
  \color{#2}{%
  }\bfseries{}##1}}%
  \MakeRobustCommand#1}

\makeauthor{\lk}{purple}
\makeauthorNew{\Nlk}{purple}
\makeauthor{\af}{orange}
\makeauthor{\vc}{green}
\makeauthor{\sr}{brown}
\makeauthor{\ajm}{blue}

\makeatletter
\AtBeginDocument{\let\LS@rot\@undefined}
\makeatother

\begin{document}

\title{Theory of intervalley-coherent AFM order and topological superconductivity in \texorpdfstring{\tWSe{}}{WSe2}}

\author{Ammon Fischer}
\email{ammon.fischer@rwth-aachen.de}
\affiliation{Institute for Theory of Statistical Physics, RWTH Aachen University, and JARA Fundamentals of Future Information Technology, 52062 Aachen, Germany}
\affiliation{Max Planck Institute for the Structure and Dynamics of Matter, Center for Free Electron Laser Science, 22761 Hamburg, Germany}

\author{Lennart Klebl}
\affiliation{Institute for Theoretical Physics and Astrophysics
and Würzburg-Dresden Cluster of Excellence ct.qmat,
University of Würzburg, 97074 Würzburg, Germany}
\affiliation{I. Institute of Theoretical Physics, Universität Hamburg, Notkestraße 9-11, 22607 Hamburg, Germany}

\author{Valentin Cr\'epel}
\affiliation{Center for Computational Quantum Physics, Flatiron Institute, New York, NY 10010, USA}

\author{Siheon Ryee}
\affiliation{I. Institute of Theoretical Physics, Universität Hamburg, Notkestraße 9-11, 22607 Hamburg, Germany}
\affiliation{The Hamburg Centre for Ultrafast Imaging, 22761 Hamburg, Germany}

\author{Angel Rubio}
\affiliation{Max Planck Institute for the Structure and Dynamics of Matter, Center for Free Electron Laser Science, 22761 Hamburg, Germany}
\affiliation{Center for Computational Quantum Physics, Simons Foundation Flatiron Institute, New York, NY 10010 USA}

\author{Lede Xian}
\affiliation{Tsientang Institute for Advanced Study, Zhejiang 310024, China}
\affiliation{Songshan Lake Materials Laboratory, 523808 Dongguan, Guangdong, China}
\affiliation{Max Planck Institute for the Structure and Dynamics of Matter, Center for Free Electron Laser Science, 22761 Hamburg, Germany}

\author{Tim O.~Wehling}
\affiliation{I. Institute of Theoretical Physics, Universität Hamburg, Notkestraße 9-11, 22607 Hamburg, Germany}
\affiliation{The Hamburg Centre for Ultrafast Imaging, 22761 Hamburg, Germany}

\author{Antoine Georges}
\affiliation{Coll\`ege de France, 11 place Marcelin Berthelot, 75005 Paris, France}
\affiliation{Center for Computational Quantum Physics, Flatiron Institute, New York, NY 10010, USA}
\affiliation{CPHT, CNRS, Ecole Polytechnique, Institut Polytechnique de Paris, Route de Saclay, 91128 Palaiseau, France}
\affiliation{DQMP, Universit{\'e} de Gen{\`e}ve, 24 quai Ernest Ansermet, CH-1211 Gen{\`e}ve, Suisse}

\author{Dante M.~Kennes}
\affiliation{Institute for Theory of Statistical Physics, RWTH Aachen University, and JARA Fundamentals of Future Information Technology, 52062 Aachen, Germany}
\affiliation{Max Planck Institute for the Structure and Dynamics of Matter, Center for Free Electron Laser Science, 22761 Hamburg, Germany}

\author{Andrew J. Millis}
\affiliation{Center for Computational Quantum Physics, Flatiron Institute, New York, NY 10010, USA}
\affiliation{Department of Physics, Columbia University, 538 West 120th Street, New York, NY 10027, USA}

\begin{abstract}
The recent observation of superconductivity in the vicinity of Fermi surface reconstructed insulating or metallic states has established twisted bilayers of WSe\textsubscript{2} as an exciting platform to study the interplay of strong electron-electron interactions, broken symmetries and topology. 
In this work, we use a first-principles, material-specific theoretical treatment that is unbiased with respect to electronic instabilities to study the emergence of electronic ordering in twisted WSe\textsubscript{2} driven by gate-screened Coulomb interactions. We construct exponentially localized moiré Wannier orbitals that faithfully capture the bandstructure and topology of the system, project the gate-screened Coulomb interaction onto them and use unbiased functional renormalization group techniques to resolve the momentum and orbital structure of the leading instabilities and the relevant energy scales. We find an interplay between intervalley-coherent antiferromagnetic (IVC-AFM) order and chiral, mixed-parity $d/p$-wave superconductivity for carrier concentrations near a displacement field and twist-angle-tunable van-Hove singularity.
Our microscopic approach establishes incommensurate IVC-AFM spin fluctuations as the dominant electronic mechanism driving the formation of superconductivity in $\theta = 5.08^{\circ}$ twisted WSe\textsubscript{2} and explains key aspects of recent experiments including the asymmetric density dependence of the spin ordering with respect to the van-Hove line, the single and double-peak structure of the DOS in the ordered (hole-doped) IVC-AFM phase, the emergence of superconductivity as the density is varied across the van-Hove line and the evolution of the displacement field-density phase diagram with twist angles between $3.7^{\circ} \dots 5^{\circ}$. 
We show that the interplay of electronic correlations and non-trivial quantum geometry in \tWSe{} manifests in orbital-selective order parameters associated to the IVC-AFM and SC states which are detectable by local spectroscopy measurements.
\end{abstract}

\maketitle

\section{Introduction}

Transition metal dichalcogenide (TMD) heterostructures have arisen as fertile and highly versatile platforms for exploring the interplay between topology and strong electronic correlations~\cite{Kennes2021}, hosting correlated insulators~\cite{wang2020correlated,xu2020correlated,huang2021correlated}, spin-ordered metals~\cite{ghiotto2024stoner,anderson2023programming,crepel2023anomalous}, rich multi-orbital physics~\cite{zhang2021electronic,ryee2023switching},
Feschbach resonances~\cite{slagle2020charge,crepel2021new,schwartz2021electrically,kuhlenkamp2022tunable,crepel2023topological,wagner2023feshbach,lange2024pairing}, quantum critical behavior~\cite{ghiotto2021quantum,li2021continuous}, excitonic insulators~\cite{tran2019evidence,ma2021strongly,gu2022dipolar,xiong2023correlated}, topological Kondo physics~\cite{dalal2021orbitally,guerci2023chiral,zhao2023gate,guerci2024topological,zhao2023emergence,xie2024kondo}, quantum anomalous and quantum spin-Hall insulators~\cite{tao2024valley,kang2024double}, ferromagnetic fractional quantum anomalous phases~\cite{zeng2023thermodynamic,cai2023signatures,xu2023observation,park2023observation}, or gapped spin liquids~\cite{kang2024evidence,crepel2024spinon}. Additionally, early experiments on $5.0^\circ$ twisted WSe\textsubscript{2}  hinted at the existence of superconducting phases in moir\'e TMDs~\cite{wang2020correlated}; a finding recently confirmed by sub-Kelvin transport measurements on $3.65^\circ$~\cite{xia2024unconventional}, $4.6^{\circ}$~\cite{xia2025simulatinghightemperaturesuperconductivitymoire} and $5.1^\circ$~\cite{guo2024superconductivity} twisted samples. 
The discovery of superconductivity in these highly tunable moir\'e TMDs has sparked many theoretical studies~\cite{schrade2021nematic,wietek2022tunable,klebl2023competition,kim2024theory,zhu2024theory,christos2024approximate,guerci2024topologicalSC,tuo2024theorytopo,chubukov2024quantum,xie2024superconductivity,qin2024kohn,chubukov2024quantum} aimed at understanding the origin and properties of strongly correlated topological superconductivity. Exciting proposals for trapping non-abelian anyons at clean, gate-defined interfaces have also appeared~\cite{mong2014universal,vaezi2014superconducting,barkeshli2016charge,crepel2024attractive}.

The association of superconductivity with proximity to the non-integer carrier concentrations at which the Fermi surface contains a van-Hove singularity (VHS), and the proximity to an apparent metallic antiferromagnetic phase suggests a weak-to-moderate coupling scenario in the $5.1^\circ$ device, where the effects of electron-electron interactions are amplified by the density of states divergence associated with the van-Hove point. The association of superconductivity at particular values of the displacement field may then be related to proximity to a higher order van-Hove singularity (HOVHS) with a stronger density of states divergence  appearing as the band theory is tuned~\cite{bi2021excitonic}. Recent literature has presented various versions of this scenario. Interactions considered include spin-fluctuations~\cite{klebl2023competition,schrade2021nematic,chubukov2024quantum}, over-screened Coulomb interactions~\cite{guerci2024topologicalSC} and phenomenological attraction~\cite{zhu2024theory,tuo2024theorytopo}. While these treatments provide important insights, they are based on particular {\it ans\"atze} and come to different conclusions regarding the symmetry of the superconducting state.  

Moreover, theory should take into account the non-trivial topological properties~\cite{devakul2021magic,crepel2024bridging,Zhang_2024} of the low-lying hole bands. In addition to potentially endowing the superconducting state with interesting properties~\cite{guerci2024topologicalSC}, they obstruct an effective single-orbital description of the topmost moiré valence band and imply that interactions projected onto the top-most band generically have a long-ranged structure. 
A minimal (topologically unobstructed) faithful low-energy model of \tWSe{} comprises three orbitals per valley that are centered on the triangular and honeycomb sites of the moir\'e superlattice~\cite{crepel2024bridging,qiu2023}. Such a model  accurately accounts for the topological entanglement of the topmost valence bands with Bloch states far below the Fermi energy and for the subtle aspects of the fermiology and its dependence on the displacement field. In this multi-orbital Wannier basis, the gate-screened Coulomb interactions relevant for the experimental setup can be cast into fairly local interactions, which ensures the applicability of established many-body techniques beyond the mean-field level~\cite{tuo2024theorytopo,guerci2024topologicalSC}, including the functional renormalization group~\cite{salmhofer2001fermionic, metzner2012functional, platt2013functional, dupuis2021nonperturbative, beyer2022reference} (FRG), dynamical mean-field theory~\cite{georges1996dynamical}, or tensor network based approaches~\cite{schollwock2011density,orus2014practical}. A similar situation arises in twisted bilayers of graphene, where a multi-band ``heavy fermion'' description is argued to best represent the physics~\cite{carr2019wannier,song2022magic}.
%

%%%%%%%%%%%%%%%
In this work, we go beyond previous treatments of correlated phases in \tWSe{} and provide a first-principles, theoretically unbiased analysis of correlated states in the weak-to-moderate interaction regime. Our theory is based on  physically realistic repulsive gate-screened Coulomb interactions projected onto a multi-band model based on exponentially localized Wannier functions that capture the quantum geometry and the intricacies of the band structure.
Using functional renormalization group techniques that are unbiased in the sense of not a priori favoring any particular electronic instability, we map out the interacting phase diagram of \tWSe{} in the twist angle range $\theta = 3.7^{\circ} \dots 5^{\circ}$ from \emph{first principles}. 
We demonstrate that intervalley-coherent anti-ferromagnetic (IVC-AFM) order and extended regions of chiral $d/p$-wave superconductivity coexist along the gate-tunable van-Hove line. 
The  unbiased characterization of particle-hole and particle-particle instabilities as well as their momentum-agnostic properties in conjunction with the subtle fermiology near the van-Hove line whose analysis is enabled by our choice of methods permits us to  show that the electronic driving mechanism for superconductivity in \tWSe{} is primarily facilitated by spin fluctuations around regions of incommensurate IVC-AFM order below the higher order van-Hove singularity, where the absence of perfect nesting strengthens particle-particle fluctuations.
The topology of the system manifests in orbital-selective order parameters that have weight only in certain domains of the \tWSe{} superlattice.
The interplay of electronic correlations and the non-trivial quantum geometry of the moiré flat bands manifests in orbital-selective order parameters that have weight only in certain domains of the \tWSe{} superlattice.
Our microscopic approach can explain the most pivotal features observed in recent experimental measurements. For the the twist angle of $\theta=5.08^{\circ}$ this includes the unusual density dependence of the particle-hole instabilities with respect to the van-Hove line, the transition from a single-peak to a double-peak structure in the density of states (DOS) as well as the subtle interplay of IVC-AFM order and superconductivity along the tunable van-Hove line.
As the twist angle is decreased below $\theta=5.08^{\circ}$, our theory predicts a successive shift of IVC-AFM domains towards the commensurate filling $n/n_0 \sim -1$ and smaller values of the displacement field accompanied by an increasing asymmetry of particle-hole instabilities with respect to the van-Hove line.
The superconducting domains in the phase diagram are eventually displaced from the van-Hove line. The accompanying decrease in the density of states is found to be more important than the increase in interaction, leading to a reduction in critical temperature.
Our theoretical analysis indicates that the leading SC order parameter is of mixed-parity, chiral $d/p$-wave character, for all twist angles studied, however, its orbital decomposition changes such that the system transitions from a multi-orbital to a single-orbital superconductor.

This paper is structured as follows: In Section~\ref{sec:review} we present the basic physics of the system. In Section~\ref{sec:frg}, we describe how to construct a faithful,  multi-orbital Wannier model that captures the intricate fermiology and quantum geometry of \tWSe{} and accounts for the dual-gated Coulomb interaction between the electrons and discuss our treatment of electronic correlations within the framework of the functional renormalization group (FRG) that allows an \textit{unbiased, beyond mean-field} characterization of emergent phases in \tWSe{}. In Section~\ref{sec:frg}, we map out the FRG phase diagram of $\theta=5.08^{\circ}$ and unravel a delicate interplay between IVC-AFM order and chiral $d/p$-wave superconductivity along the displacement-field tunable van-Hove line. In Section~\ref{sec:ivc-afm}, we first provide a detailed characterization of the IVC-AFM state and access quasiparticle properties by performing a FRG+MF analysis in the ordered phase. In particular, we demonstrate how significant electronic screening contributions accumulated by particle-particle scattering within FRG leads to a suppression of effective Hubbard-like interactions, which bridges to the observed over-estimation of electronic order observed in mean-field (Hartree-Fock) approaches.
In Section~\ref{sec:sc}, we discuss the electronic mechanism leading to the formation of chiral $d/p$-wave superconductivity in \tWSe{} and its relation to the nearby IVC-AFM domains in the phase diagram. We provide an in-depth analysis of spectral properties in the SC phase and unravel a strong orbital-selectivity that manifests in nodal signatures in the local density of states (LDOS) and is related to the topological properties of the Wannier model. In Section~\ref{sec:long}, we provide a detailed analysis on the role of long-ranged density-density interactions in the formation of IVC-AFM and SC order. Having established our methodology for $\theta=5.08^{\circ}$ \tWSe{}, we finally map out the twist angle evolution of IVC-AFM and SC order in the range of $\theta = 3.7 \ldots 5^{\circ}$ in Section~\ref{sec:twist}.

\begin{figure*}[t!]
    \centering
    \includegraphics[width=\textwidth]{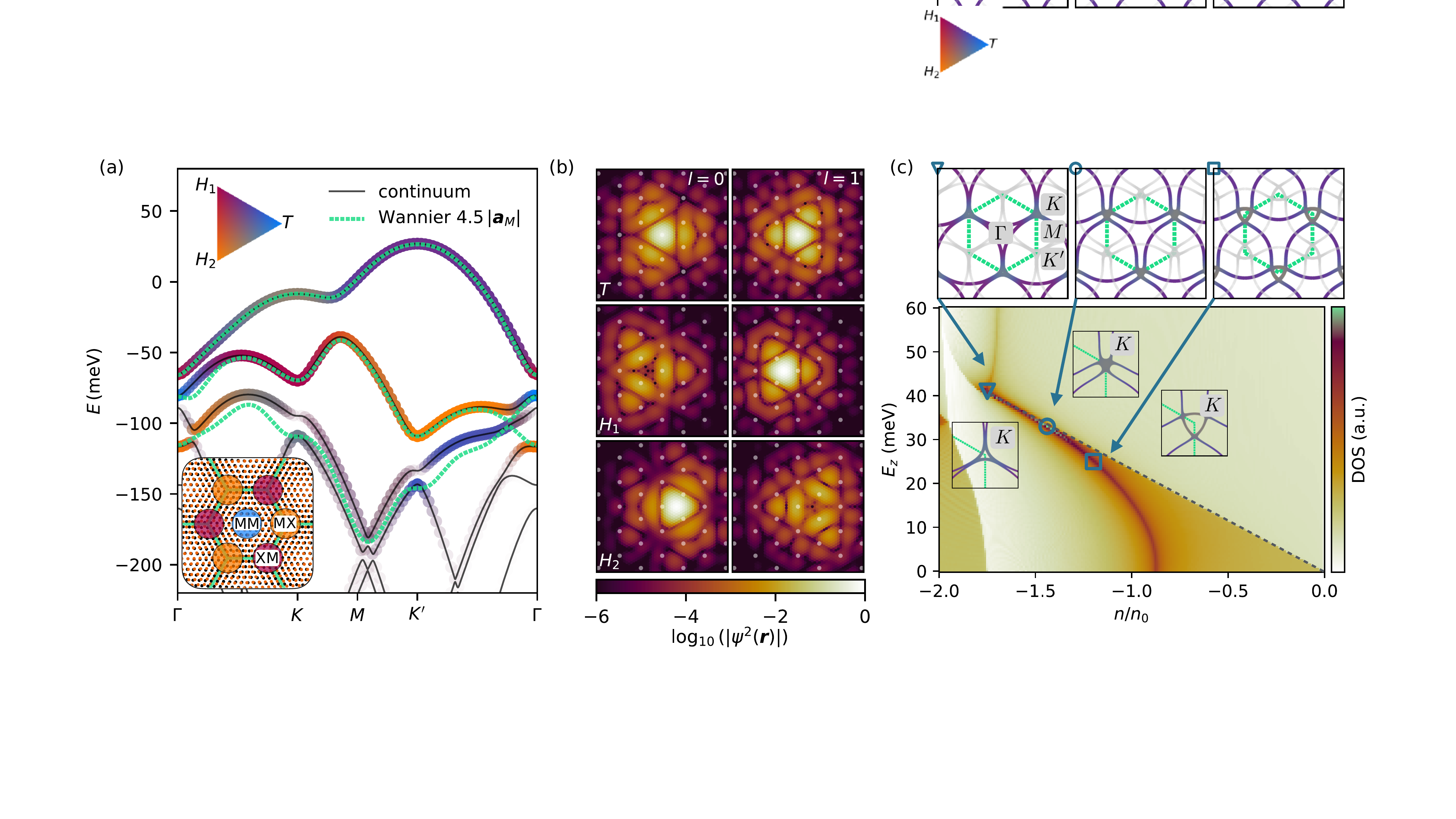}
    \caption{%
    %\textbf{
    Fermiology and Wannierization of $\theta = 5.08^{\circ}$ \tWSe{} for an external displacement field $E_z=20\,\mathrm{meV}$.
    %}
    %
    (a)~Continuum bandstructure in valley $\nu=+1$ (black) and spectral overlap of its Bloch states with a three-orbital model comprising Wannier orbitals centered at the MM stacking regions ($T$, $s$-orbital) and XM/MX stacking regions ($H_1$, $p_+$-orbital)/($H_2$, $p_+$-orbital) of the moiré superlattice (tri-colored map and inset).
    The bandstructure of the effective Wannier model truncated after 4.5 moir\'e lattice vectors is indicated by the green dashed line.
    (b)~Amplitude $|\psi_l\bvec (r)|^2$ of the Wannier functions ($T, H_1, H_2$) on the two layers ($l=0,1$) that capture the spectral weight of the continuum model.
    (c)~Density of states as function of holes per moiré unit cell $n/n_0$ and displacement field $E_z$. 
    The upper panel shows snapshots of the Fermi surface along the van-Hove line and its orbital polarization using the same colormap as in (a) for valley $\nu=+1$. The Fermi surface of the opposite valley $\nu=-1$ is shown in gray. The green dashed line indicates the mini-Brillouin zone of \tWSe{}.
    The blue circle indicates the position of the higher-order VHS, where three van-Hove points merge at $K^{\nu}$ as demonstrated in the insets. The three van-Hove points are moved towards $\Gamma$ ($M$) for increasing (decreasing) value of the external displacement field along the van-Hove line as indicated by the blue triangle (square). 
    }
    \label{fig:wannier}
\end{figure*}

\section{\texorpdfstring{WSe\textsubscript{2}}{WSe2m} overview}\label{sec:review}

Twisted WSe\textsubscript{2} (\tWSe{}) consists of two monolayers of WSe\textsubscript{2} stacked at a relative twist angle $\theta$ and encapsulated in a structure including top and bottom gates. 
The twist, in combination with the electrostatic potential from the local arrangement of WSe\textsubscript{2} bilayers and the interlayer electronic hybridization, produces a large unit cell with two dimensional  ``moir\'e''  bands~\cite{wu2019topological,devakul2021magic}. The average of the potentials on the top and bottom gates tunes the electronic density over wide ranges. 
The difference in voltage between the top and bottom gates introduces an interlayer potential difference, or ``displacement field'', which shifts electronic wave functions and charge density between the layers, modifying the electronic band structure. At zero and small displacement fields and hole concentrations not too different from one per unit cell, the bands near the Fermi surface are derived from a complicated combination of states distributed over both layers and originating from various regions of the moiré unit cell. This complicated combination of states leads to a non-trivial Fermi surface shown in Fig.~\ref{fig:wannier}(b) and non-trivial quantum geometry that (as will be seen) leads to a nontrivial Chern number of the superconducting state. Increasing the displacement field at fixed carrier density drives the system into a "layer-polarized" regime in which the near Fermi surface states are drawn primarily from one layer and one region of the moir\'e unit cell. In this regime, superconductivity is absent and the quantum geometry of the bands is simple.
At a fixed twist angle, the system's emergent behavior can be explored in a phase diagram defined by electronic density and displacement field. As the density varies at a fixed displacement field, the Fermi surface can pass through a van-Hove singularity (VHS), where the density of states (DOS) diverges. Of particular relevance for this study is that both the density at which a van-Hove singularity occurs and the strength of the resulting divergence in the DOS depend on twist angle and are tunable over wide ranges by varying the displacement field.

In the $5.1^\circ$ \tWSe{} device~\cite{guo2024superconductivity}, the superconducting state is observed near hole densities $n/n_0 \sim n_{\text{VHS}} \neq -1$ at which the Fermi surface passes through the van-Hove singularity, and only for a small range of displacement fields. The superconducting phase is adjacent to a metallic region characterized by an increase in longitudinal resistance and a very rapid non-monotonic doping dependence of the Hall resistivity. This unusual metallic region  has been interpreted \cite{guo2024superconductivity} as arising from the emergence of antiferromagnetic (AFM) spin ordering that partially gaps out the Fermi surface.
The densities at which superconductivity and antiferromagnetism occur are not close to one hole per moir\'e unit cell and the experimentally important displacement fields are significant, on the order of half of the field required to enter the layer-polarized regime. As the twist angle is decreased, the density range where superconductivity and antiferromagnetism are observed moves closer to one hole per unit cell and the relevant displacement field decreases. In the $3.65^\circ$ device~\cite{xia2024unconventional}, the antiferromagnetic phase is centered at hole densities of $n/n_0=-1$ per moir\'e unit cell and is insulating, presumably because at this commensurate carrier concentration the commensurate antiferromagnetic order can fully gap the Fermi surface. The antiferromagnetic insulating phase exists for displacement fields between a small but non-zero value, and the ``layer-polarization" displacement field $E_Z^{\text{lp}}$ beyond which the system enters the layer polarization regime. In this device superconductivity is also found only for carrier concentrations near $n/n_0=-1$ per moir\'e unit cell and small displacement fields adjacent to the insulating state. 

In the next Section we present our theoretical model, which encodes the twist angle-dependent fermiology and the relevant interacting physics.

\section{Model and method} \label{sec:method}
\subsection{Multi-orbital Wannier model}
The non-interacting electronic structure of hole-doped \tWSe{} is accurately described by a continuum model in which the spin-valley locked holes of both monolayers are treated within the effective mass approximation, and are coupled by moir\'e potentials that slowly vary on the atomic scale~\cite{wu2019topological}. 
Due to the emergent spin-valley locking in monolayer WSe$_2$, the low-energy states can be assigned a well-defined valley index $\nu=\pm 1$ and states from opposite spin-valley sectors are related by virtue of time-reversal symmetry $\mathcal{T}$. 
The effects of out-of-plane electric fields are captured by a layer-dependent potential energy $\pm E_z$. 
For twist angles $\theta \gtrsim 3^\circ$, in-plane lattice relaxation is weak, and the associated moir\'e fields are characterized by only a few harmonics~\cite{carr2018relaxation, zhang2024polarization, jia2024moire, thompson2024visualizing, liu2024imaging}. Here, we use the continuum model from Ref.~\cite{devakul2021magic}, whose coefficients have been fitted to large-scale \textit{ab-initio} calculations, see Supplementary Materials (SM)~\cite{SM} for details.

As most many-body methods require a symmetry-preserving lattice description of the system under scrutiny, we seek to construct a Wannier representation of the topmost valence band states in each valley based on the three-orbital model put forward in Refs.~\cite{crepel2024bridging,qiu2023}. 
This model is based on generic symmetry arguments and comprises one triangular lattice site $T$ in the MM stacking regions of the triangular superlattice (Wyckoff position 1a), which has weight on both layers, as well as two honeycomb lattice sites $H_{1,2}$ in the XM/MX stacking regions (Wyckoff positions 2a,b) with dominant weight on either one of the layers.
At a twist angle of $\theta=5.08^\circ$, however, the Chern number sequence of the first three valence bands of \tWSe{} in valley $\nu = \pm 1$ is $(\nu, \nu, \nu)$ for the parameters of Ref.~\cite{devakul2021magic}, prohibiting an exponential localization of Wannier functions centered at aforementioned Wyckoff positions constructed from the three topmost bands alone. 
We resolve this issue by employing a single-shot Wannierization scheme~\cite{marzari2012maximally, carr2019derivation, fischer2024supercell} in which Bloch states from the entire valence band manifold are considered for the projection and a proper subspace selection is instead achieved by weighting Bloch states according to their energetic hierarchy and their weight around the desired orbital center. 
Compared to the layer-polarization scheme presented in Ref.~\cite{crepel2024bridging}, the energetic subspace selection ensures to construct exponentially localized Wannier functions irrespective of the Chern number of the third moir\'e band, which sensitively depends on model parameters due to the large energy overlap of this band with remote energy bands.
\Cref{fig:wannier}~(a) shows the band structure of the Wannier model (dashed cyan lines) compared to the continuum model (black lines) at a displacement field $E_z=20\,\mathrm{meV}$. By adding the overlap of each Wannier function with the continuum model's Bloch states as color, we demonstrate how states from the fourth and fifth bands are essential for a three-orbital description with an overall Chern number sequence $(\nu,\nu,-2\nu)$. 
The residual spectral weight in remote energy bands is required to render the entire Wannier band manifold topologically trivial. In spite of this, the topmost band of the continuum model is perfectly reproduced in the Wannier model as shown in \cref{fig:wannier}~(a).

In accordance with previous studies, the topmost band has significant weight in the $T$-orbital~\cite{wang2020correlated, pan2020band, zang2022dynamical, kiese2022tmds, klebl2023competition, ryee2023switching, zegrodnik2023mixed, motruk2023kagome, wang2023staggered}, but the nontrivial topology invalidates a description as a one-band model. This contrasts MoTe\textsubscript{2} bilayers~\cite{crepel2024bridging,qiu2023} in which more weight lies in the $H_{1,2}$ orbitals, leading to different physical behaviors. 
%which could be the source of the different physical behaviors of these twisted heterostructures. 
In \cref{fig:wannier}~(b), we display the real-space structure of the three Wannier orbitals. 
Their exponential localization implies short-ranged hopping parameters as demonstrated in \cref{fig:wannier}~(a): Disregarding hopping parameters for distances $d \gtrsim 4.5\,|\bvec a_M|$ (with $\bvec a_M$ the moir\'e lattice vectors) does not change the resulting band structure significantly. A detailed analysis of the energetic and topological properties of the Wannier functions is provided in the SM~\cite{SM}.
We carry out the Wannierization for several values of displacement field $E_z$ and interpolate the hopping parameters in order to obtain a Hamiltonian for arbitrary $E_z$. The resulting density of states is a smooth function of $E_z$ [see \cref{fig:wannier}~(c)], and displays a clear maximum when the Fermi energy meets the VHS of the topmost valence band~\cite{schrade2021nematic}. Along this line, an exceptional point occurs at $E^c_z\approx 33\,\mathrm{meV}$ (blue circle), where three van-Hove points merge with each other and form a HOVHS~\cite{bi2021excitonic}. 
For $E_z < E_z^c$, the three van-Hove points move towards the $M$-point forming small triangular Fermi pockets around $K^{\nu}$ that are exclusively $T$/$H_2$-polarized. In contrast, for $E_z > E_z^c$, the three van-Hove points move towards $\Gamma$, and the Fermi pockets around $K^{\nu}$ disappear. The remaining dispersive part of the Fermi surface is entirely $T$/$H_2$-polarized for all values of the displacement field. 
Another exceptional line (grey dashed line) in the phase diagram segregates two regimes in which the $H_2$-orbital is either depleted or filled with holes. In the former regime at low densities, holes exclusively occupy the $T/H_1$-orbitals leading to a sharp drop in the DOS when decreasing the hole doping $n/n_0$. 
In previous works~\cite{guo2024superconductivity}, this regime has been dubbed 'layer-polarized' regime, however, we note that this terminology is in fact inaccurate owing to the partial occupancy of the $T$-orbital that has spectral weight in both layers as shown in Fig.~\ref{fig:wannier}~(b).
The multi-orbital Wannier model has the symmetry group $\mathcal{G}_0 = C_{3z} \otimes U_{\nu}(1) \otimes \mathcal{T}$, where $C_{3z}$ is the point group of the moiré unit cell that contains a three-fold rotation around the $z$-axis, the $U_{\nu}(1)$-symmetry acts in spin-valley space and $\mathcal{T}$ denotes time-reversal symmetry. Due to the absence of cubic terms to lowest order in the continuum description~\cite{crepel2024bridging}, the system we study theoretically further possesses a mirror-symmetry in each valley $\mathcal{M}_x$ that is weakly broken in the physical bilayer but is visible in the fermiology of a single spin-valley sector in \cref{fig:wannier}~(c). 

\subsection{Dual-gated Coulomb interactions}
To model the long-ranged Coulomb interaction between electrons in \tWSe{}, we employ an experimentally motivated interaction profile $V(r)$ that accounts for the dual-gated device architecture as well as short-ranged interactions that are regularized by an Ohno form~\cite{throckmorton2012fermions,fischer2024supercell}:
\begin{equation}
    V(r) = 4V_0\sum_{k=0}^\infty K_0\left[ (2k+1)\pi\,\frac{\sqrt{r^2+a^2}}{\xi} \right] \,.
\label{eq:dual-gated-coulomb}
\end{equation}
Here, $K_0$ denotes a modified Bessel function of the second kind, $\xi$ is the distance from the sample to either gate and $\alpha = 14.40$\,eV \AA{} is the fine-structure constant. The interaction parameters $a = \alpha/(\epsilon U)$ and $V_0=\alpha/(\epsilon\xi)$, only depend on the dielectric constant $\epsilon$ and the on-site interaction strength $U$%
%momentum average of the band projected interaction in the parabolic portion of the WSe\textsubscript2 valence bands $U$
, see SM~\cite{SM} for details. Motivated by the experimental gating architecture~\cite{xia2024unconventional,guo2024superconductivity}, we choose $\xi = 100$ \AA{} and treat the exact value of the dielectric constant as parameter as it further depends on internal screening in \tWSe{} mediated by remote bands and external screening by the hBN encapsulating layers. In the main text, we choose $\epsilon = 16$ in agreement with first-principle estimates~\cite{laturia2018dielectric}, however, provide additional data for reasonable values of $\epsilon$ in the SM~\cite{SM} that demonstrate the robustness of our results.

The dual-gated Coulomb interaction defined in Eq.~\eqref{eq:dual-gated-coulomb} is then projected onto the multi-orbital Wannier model, giving rise to an effective low-energy Hamiltonian of the form
\begin{equation}
\begin{multlined}
H=\sum_{\nu,\bvec R,\bvec R'}\sum_{X,X'} t^{\nu}_{\bvec R X,\bvec R'X'}c^{\nu\,\dagger}_{\bvec R X} c^{\nu\vdagger}_{\bvec R' X'}  \\
+ \sum_{\nu,\bvec R}\,\sum_{X\in\{T, H_1,H_2\}} \frac{U_X}2 n_{\bvec RX}^\nu n_{\bvec RX}^{\bar\nu}
\,,
\end{multlined}
\label{eq:fullham}
\end{equation}
where $\nu$ denotes the locked spin-valley degree of freedom, $X,X'$ orbital, and $\bvec R, \bvec R'$ Bravais lattice vectors. The operator $c_{\bvec RX}^{\nu\,(\dagger)}$ destroys (creates) an electron with spin/valley $\nu$ in the orbital $X$ on site $\bvec R$, and $n^\nu_{\bvec RX} = c^{\nu\,\dagger}_{\bvec RX} c^{\nu\vdagger}_{\bvec RX}$ is the density operator. As the hopping parameters $t_{\bvec RX,\bvec R'X'}^{\nu}$ are valley-diagonal, the interaction part is what couples different spins/valleys. To disentangle the microscopic mechanisms responsible for the formation of the Fermi surface reconstructed AFM phase and superconductivity observed in experimental measurements~\cite{xia2024unconventional,guo2024superconductivity}, we first consider the dominant on-site interaction associated to the triangular ($U_T$) and honeycomb ($U_{H_{1,2}}$) sites and later investigate the influence of long-ranged density-density interaction on the phase diagram of $\theta=5.08^{\circ}$ \tWSe{}. We further provide a detailed list of the effective interaction parameters in the Wannier model for different values of $(\xi, \epsilon)$ in the SM~\cite{SM}.
The values of $U_T \approx 74\,\mathrm{meV}$ and $U_{H} \approx 106\,\mathrm{meV}$ place \tWSe{} into the weak-to-intermediate coupling regime, where the band width $W \approx 200\,\mathrm{meV}$ of the three-orbital model fulfills $W \gtrsim U_{T/H}$. We note that due to the topological entanglement of the topmost valence bands in \tWSe{}, the bandwidth of the entire Wannier band manifold (and not only the topmost valence band) should be considered.

\begin{figure*}
    \centering
    \includegraphics[width=1\textwidth]{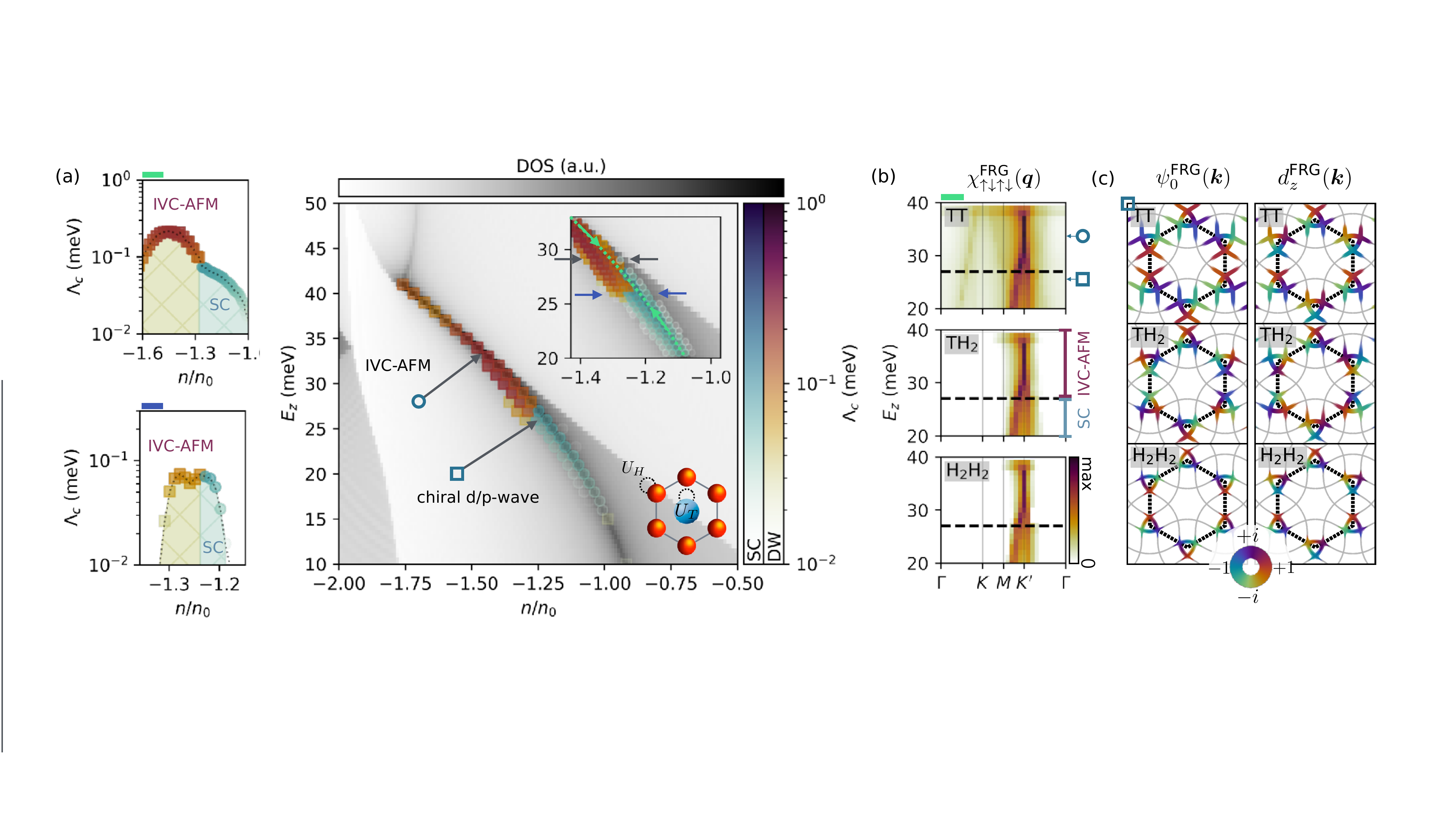}
    \caption{%
    FRG phase diagram of $\theta = 5.08^{\circ}$ \tWSe{} resolving the interplay of IVC-AFM order and chiral $d/p$-wave superconductivity along the gate-tunable van-Hove singularity.
    (a)~Main panel: FRG phase diagram as function of holes per moiré unit cell $n/n_0$ and external displacement field $E_z$. 
    The phase diagram shows the critical scale $\Lambda_{\mathrm c}$ of the leading Fermi surface instability and indicates regimes of spin/charge density wave order (DW, red) and superconductivity (SC, blue). 
    The density of states (DOS) is shown in the background (gray) to pinpoint the position of the van-Hove line.
    Close to the HOVHS (blue circle), the system features IVC-AFM order, while for lower values of the displacement field chiral $d/p$-wave superconductivity emerges on the van-Hove line.
    Inset: Zoom-in on the density-displacement field region where IVC-AFM order gives way to superconductivity.
    The colored arrows indicate different directions when moving along the van-Hove line (green arrows) or at constant displacement field cuts (grey, blue arrows). 
    The critical scale $\Lambda_{\mathrm{c}}$ along these line cuts is shown in the panels on the left, indicating a maximal critical temperature of $\Lambda_{\mathrm{c}} \sim 500\,\mathrm{mK}$ that is reached in the immediate vicinity of the IVC-AFM phase.
    (b)~Momentum structure and (in)commensurability  of the IVC-AFM state along the van-Hove line. The subpanels show the absolute value of the inter-valley particle-hole susceptibility $\chi^{\text{FRG}}_{\uparrow \downarrow \uparrow \downarrow}(\bvec q)$ at scale $\Lambda_{\mathrm{c}}$ for the relevant orbital components involving the Wannier orbitals $T/H_2$. The leading momentum transfer $\bvec Q_C$ locks to $K'$ at the HOVHS (blue circle) and successively shifts towards incommensurate momenta along the high symmetry path $K'$-$M$ when lowering the external displacement field along the van-Hove line (blue square). 
    (c)~Amplitude and phase dependence of the pairing instability at $E_z=25\,\mathrm{meV}$ (blue circle) decomposed into singlet $\psi_0(\bvec k)$ and triplet $d_z(\bvec k)$ components. The continuous colorbar indicates the relative phase, whereas the amplitude is encoded by the opacity. The SC order parameter is two-fold degenerate and likely minimizes its free energy by the chiral superposition $d+id/p+ip$.
    }
    \label{fig:frg}
\end{figure*}

\subsection{Functional renormalization group approach}\label{sec:frg}

In the weak-to-moderate coupling regime, the functional renormalization group (FRG)~\cite{salmhofer2001fermionic, metzner2012functional, platt2013functional, dupuis2021nonperturbative, beyer2022reference} represents a well-established method to predict particle-particle and particle-hole instabilities in an unbiased manner. FRG smoothly connects the non-interacting action to the interacting one by virtue of a renormalization group (RG) scale $\Lambda = \infty \to 0$. The introduction of the RG scale leads to an infinite hierarchy of coupled differential equations for the vertex functions $\Gamma^{(2n)}$. 
In order to numerically solve the coupled nonlinear differential equations, we must truncate the hierarchy and approximate the vertex functions' dependencies on frequency and momentum. To account for the topology of the multi-orbital Wannier setup for the moiré valence bands, we here employ the static four-point truncated unity FRG (TUFRG) approximation~\cite{husemann2009efficient, lichtenstein2017high, profe2022tu} that allows to keep track of the entire orbital/band and (incommensurate) momentum dependence of the model at the expense of neglecting the frequency dependence of the vertex functions. 
The level-2 truncated FRG flow equations for the two-particle vertex function $\Gamma = \Gamma^{(4)}$ read
\begin{equation}
\partial_{\Lambda} \Gamma^{\Lambda} = \sum_\gamma \partial_{\Lambda} \gamma^{\Lambda} \,, \quad \partial_{\Lambda} \gamma^{\Lambda} =  \Gamma^{\Lambda} \circ \partial_{\Lambda} L^{\gamma, \Lambda} \circ \Gamma^{\Lambda} \,,
\label{eq-frg}
\end{equation}
where $\gamma \in \{P,D,C\}$ denotes the decomposition into two-particle reducible interaction channels that capture fluctuations accumulated in the particle-particle ($P$), direct particle-hole ($D$) and crossed particle-hole ($C$) channel. Most importantly, this allows to treat all couplings on equal footing and (at one loop order) accounts for mutual feedback between the different interaction channels that are renormalized during the flow. 
In static four-point (TU)FRG, the onset of an ordered phase is signaled by the divergence of the two-particle vertex $\Gamma^{(4)}$ at a particular (critical) scale $\Lambda_\mathrm{c}$. The leading fermionic bilinear in the divergent channel is obtained by an eigenvalue decomposition of the channel-specific vertex $\gamma$
\begin{equation}
\gamma_{\kappa \kappa'}(\bvec q_\gamma) = \sum_i \phi^L_{\kappa,i}(\bvec q_\gamma) \, \gamma_i(\bvec q_\gamma) \, \phi^R_{\kappa',i}(\bvec q_\gamma) \,,
\label{eq-fermionic-bilinear}
\end{equation}
where $\kappa = (\bvec k_\gamma, X, \nu_1,\nu_2)$ is a multi-index comprising the fermionic momentum variable $\bvec k_\gamma$, Wannier function index $X$, and spin/valley $\nu_{1,2}$. The left/right eigenvectors to the channels' eigenvalues $\gamma_i$ are denoted by $\phi^{L/R}_{\kappa,i}$. They provide insight into the momentum, spin/valley and orbital structure of the order parameter. 
In the absence of spin $SU(2)$-symmetry, particle-hole instabilities in the $C/D$-channel are allowed to admix and we therefore refer to them as density wave instabilities (DW) in the following.
Since we choose the sharp frequency cutoff as regulator (as implemented in the divERGe library~\cite{profe2024diverge}), the critical scale $\Lambda_c$ serves as a proxy for the critical temperature of the transition. 
To render the solution of Eq.~\eqref{eq-frg} numerically feasible, the vertex functions are parametrized in a hybrid real-momentum space representation~\cite{profe2022tu,profe2024diverge} using a $30 \times 30$ momentum mesh for the bosonic momentum variable ($\bvec q_\gamma$) in each two-particle reducible interaction channel---hence allowing to resolve (in)commensurate particle-particle and particle-hole instabilities---whereas the fermionic momenta ($\bvec k_\gamma$) are expanded in a real-space form factor basis that is truncated after a distance of $2 \ldots 4|\bvec a_M|$. The two-particle propagator $L^{\gamma, \Lambda}$ is resolved on a refined $900 \times 900$ momentum mesh to increase the energy resolution and to sufficiently resolve the fermiology of the system at low scales. A detailed overview of the level-2 truncated FRG flow equations and technical details are provided in the SM~\cite{SM}. 

The level-2 truncation of the FRG flow adapted in Eq.~\eqref{eq-frg} breaks down when approaching the symmetry-broken phase as indicated by a divergence of the two-particle vertex $\Gamma^{\Lambda}$ at scale $\Lambda = \Lambda_{\mathrm{c}}$. To access quasiparticle properties and spectral functions in the ordered phase for scales $\Lambda_{\text{MF}} \lesssim \Lambda_{\mathrm{c}}$ predicted by FRG, we may formulate an effective mean-field (MF) theory as outlined in Ref.~\cite{metzner2012functional,wang2014competing}. To overcome the non-analytic behavior of the FRG flow at the critical scale $\Lambda_{\mathrm{c}}$ and connect the symmetry-broken state with the high-temperature normal-state, we construct a suitable bare interaction $V^{\text{MF}}$ for MF treatment by solving an inverse Bethe-Salpeter equation in the divergent channel at scale $\Lambda_{\mathrm{c}}$
\begin{equation}
    \gamma = V^{\text{MF}} + V^{\text{MF}} \circ \chi^{0,\gamma} \circ \gamma \,,
\label{eq:inv-bse}
\end{equation}
where $\chi^{0,\gamma}$ is the bare particle-particle (particle-hole) susceptibility, i.e $\chi^{0,\gamma} = \int^{\infty}_{\Lambda_{\mathrm{c}}} \text{d} \Lambda \, \partial_{\Lambda}L^{\gamma, \Lambda}$. Inverting this equation yields the bare interaction $V^{\text{MF}}$ that by construction yields the correct MF state upon a successive self-consistent MF decoupling.
We further note that the single-channel RPA approximation can be recovered from the full FRG flow equation defined in Eq.~\eqref{eq-frg} by neglecting inter-channel coupling, i.e. only considering the single-channel ladder resummation and integrating the second term in Eq.~\eqref{eq-frg} 
\begin{equation}
\gamma^{\text{RPA}} =  \Gamma^0 \circ \chi^{0,\gamma} \circ \Gamma^0 \,,
\label{eq:rpa}
\end{equation}
where $\Gamma^0$ refers to the initial (bare) interaction. 

\section{FRG phase diagram}

We present the FRG phase diagram of $5.08^\circ$ \tWSe{} in \cref{fig:frg}~(a), where we plot the critical scale $\Lambda_\mathrm{c}$ of emergent particle-particle (SC, blue) and particle-hole (DW, red) instabilities as a function of hole filling per moir\'e unit cell $n/n_0$ and transverse electrical field $E_z$. The density of states (DOS) is shown as a reference to pinpoint the position of the van-Hove line (gray). At the position of the HOVHS, the system promotes an inter-valley coherent (IVC) anti-ferromagnetic spin-density wave with order parameter
\begin{equation}
\vec \Delta^{\text{AFM}}_{\bvec R} \propto \!\!\! \sum_{X \in \{T, H_2 \}} e^{i [\bvec Q_C \cdot \bvec R + i\vartheta_X]} \langle c^{\nu\,\dagger}_{\bvec R X} \vec{\sigma}^{\nu \nu'} c^{\nu' \vdagger}_{\bvec R X} \rangle + \text{h.c.} \,,
\label{eq:sdw_order_parameter}
\end{equation}
where $\vec{\sigma} = (\sigma_x, \sigma_y)^{\mathrm{T}}$ acts in the space of the locked spin-valley degree of freedom and $\vartheta_X$ measures the relative orientation and weight of the $120^{\circ}$ in-plane spin orientation on the orbitals $T/H_2$ that is otherwise arbitrary due to the remaining $U_{\nu}(1)$ symmetry.
%is an expression of the remaining spin-$U(1)$ symmetry that measures the angle of the $120^{\circ}$ in-plane spin orientation with respect to the $\bvec \hat{e}_x$-axis. 
The spin density wave ($\mathcal{T}$-odd) breaks translational invariance due to the finite transfer momentum $\bvec Q_C = K^{\nu}$ that connects the HOVHSs in different spin-valley sectors and transforms in the trivial $A$ irreducible representation (IR) of the $C_{3z}$ little group at $K^{\nu}$. Consistent with the orbital polarization of the Fermi pockets at the HOVHS, the order parameter has dominant weight on the $T/H_2$-orbitals as imprinted in the inter-valley particle-hole susceptibility $\chi_{\uparrow \downarrow \uparrow \downarrow}^{\text{FRG}}(\bvec q)$ (see SM~\cite{SM} for details)
shown in \cref{fig:frg}~(b). We further provide a sketch of the real-space spin pattern in Fig.~\ref{fig:ivc_spectral}~(a).
Decreasing the value of the displacement field $E_z$ along the van-Hove line shifts the leading momentum transfer $\bvec Q_C$ towards the $M$-point, causing regions of incommensurate IVC-AFM order as demonstrated in \cref{fig:frg}~(b). 

The absence of a perfect scattering vector that connects the van-Hove points in different spin-valley sectors and the reduction of the DOS weakens particle-hole instabilities for $E_z<E_z^c$ and instead promotes the effect of particle-particle fluctuations leading to the formation of superconductivity along the van-Hove line, see blue square in \cref{fig:frg}~(a). 
The formation of superconductivity along the van-Hove line is hence facilitated by Fermi surface warping when detuning the system from the HOVHS, reminiscent of pairing in the $t$-$t'$ square lattice Hubbard model~\cite{honerkamp2003instabilities} as will be discussed in \cref{sec:spin-fluctuations} in more detail.
The Ising-like spin-orbit coupling in \tWSe{} couples singlet- and triplet components of the superconducting order parameter
\begin{equation}
    \hat \Delta^{\text{SC}}(\bvec k) = \left [ \hat \psi_0(\bvec k) \sigma_0 + \vec{\hat d}(\bvec k) \cdot \vec \sigma \right ] i \sigma_y \,,
\label{eq:sc_singlet_triplet}
\end{equation}
where the hat-symbol encodes the orbital and spin-valley degree of freedom. We find that the leading superconducting order parameter is an inter-valley symmetric combination ($S^z=0$) of singlet and triplet components $[\psi_0(\bvec k), d_z(\bvec k)]$ that are allowed to admix due to the preserved $U_{\nu}(1)$ symmetry~\cite{schrade2021nematic}. The Fermi surface projected amplitude and phase of the respective components is shown in the right panel of \cref{fig:frg}~(c). 
Similar to the IVC-AFM order parameter, the SC order parameter only has weight on bonds including the $T/H_2$-orbital, whereas $\Delta_{H_1 X}(\bvec k) = \Delta_{X H_1}(\bvec k) = 0$ for $X \in \{T, H_1, H_2\}$, see SM~\cite{SM} for full orbital-resolved structure of the SC order parameter.
The leading order parameter transforms in the two-dimensional $E$ irreducible representation 
of the Hamiltonian's point group $\mathcal{G}_0$.
In a mean-field picture, the two degenerate order parameters are in general superimposed and the free energy is minimized by the chiral, time-reversal breaking combination $d\pm id/p\pm ip$. The latter gaps the Fermi surface and hence minimizes the energy of the superconducting condensate~\cite{honerkamp2003instabilities,nandkishore2012chiral,black-schaffer2014chiral,classen2019competing,scherer2022chiral}. The chiral superconducting state belongs to the A Cartan (Altland-Zirnbauer) class for $S_z$ preserving Bogoliubov de-Gennes Hamiltonians~\cite{schneyder2008classification} possessing a $\mathbb{Z}$ topological invariant that manifests in a non-vanishing Chern number $C= \pm 2$ and edge modes that are described by Dirac-like complex fermionic quasiparticle.

Besides the microscopic characterization of the emergent phases, a central question of recent experimental measurements concerns their interplay along the displacement field tunable van-Hove line that is captured by the unbiased FRG analysis. To this end, the inset of \cref{fig:frg}~(a) details the transition region of the phase diagram from the intervalley-coherent (IVC) antiferromagnet (AFM) to the chiral $d/p$-wave superconductor. We further provide line cuts of the phase diagram along the van-Hove line (green dash) and for constant displacement field (blue dash) in \cref{fig:frg}~(a). Further line cuts at constant displacement field are provided in the Supplementary Material~\cite{SM}. We find that the highest transition temperature of the superconducting state $\Lambda_c \sim 500\,\mathrm{mK}$ is reached in the immediate vicinity to the (incommensurate) IVC-AFM phase along the van-Hove line. 
The superconducting region flanks the IVC-AFM state and we observe an asymmetry of spin-ordered regimes with respect to the van-Hove line, i.e. instabilities occur at larger critical scales $\Lambda_c$ for $n < n_\mathrm{VHS}$ and $E_z<E_z^c$ in agreement with experimental observations~\cite{xia2024unconventional,guo2024superconductivity}. 

\section{Quasi-particle properties and asymmetry of the IVC-AFM state}\label{sec:ivc-afm}

\begin{figure}[t!]
    \centering
    \includegraphics[width=\columnwidth]{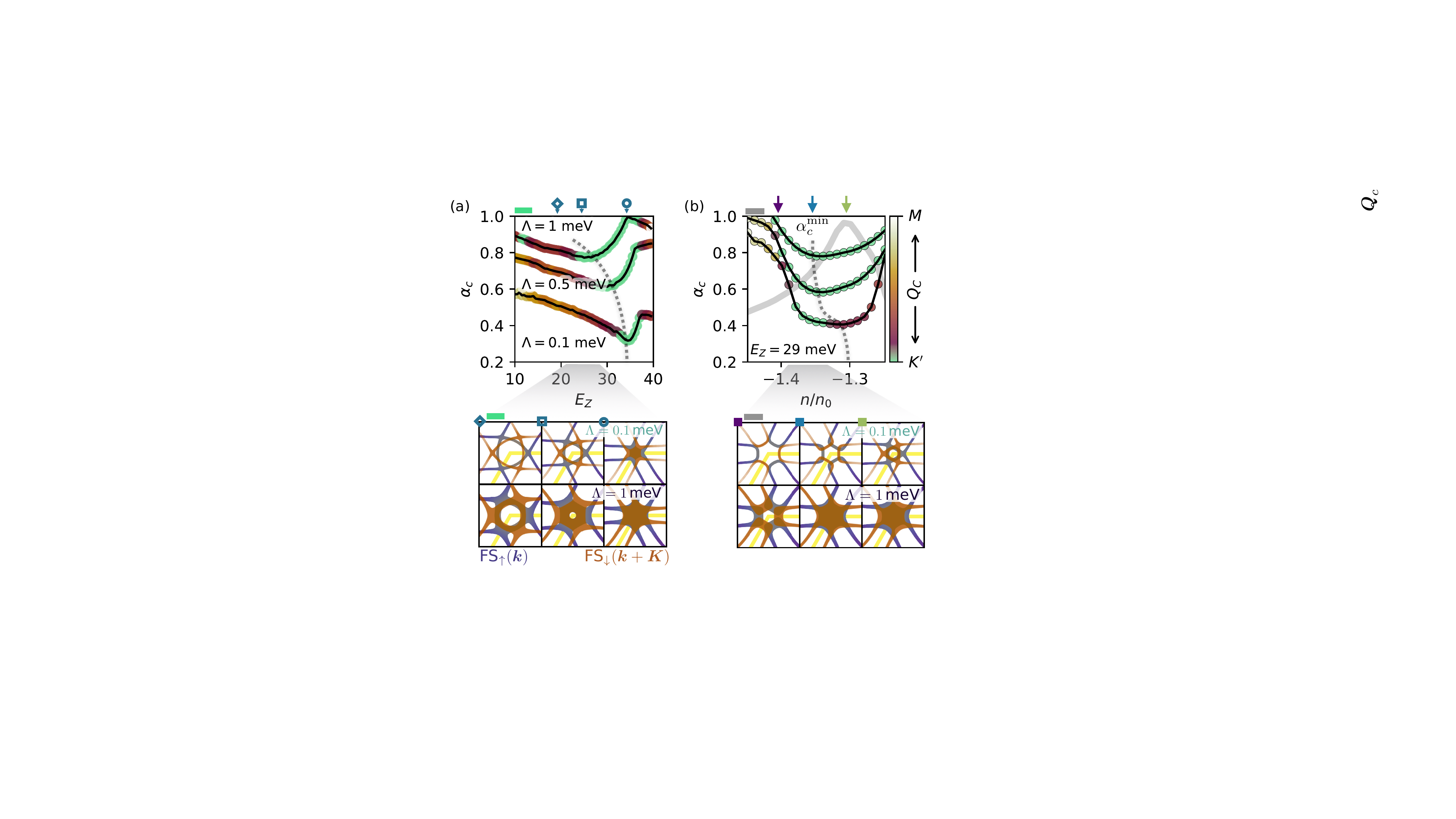}
    \caption{%
    %\textbf{
    Asymmetry of particle-hole instabilities due to Fermi surface broadening induced nesting.
    %}
    %Stuff on assymtry
    (a)~Critical (relative) interaction scale $\alpha_c$ of the RPA renormalized interaction for different scales $\Lambda$ along the van-Hove line. The leading momentum transfer $\bvec Q_C$ along the high symmetry path $K^{\nu}$-$M$ is encoded by the colorbar shared among panels~(a,b). The minimal value of $\alpha_c$ for each curve is indicated by the gray dashed line showing a clear asymmetry of particle-hole instabilities with respect to the position of the HOVHS (blue circle). 
    The lower panel shows the ($\Lambda$-broadened) Fermi surface sheets of the spin-valley polarized sub-sectors, i.e., all eigenenergies that are within the energy shell $[E_F - \Lambda, E_F + \Lambda]$ around the Fermi energy $E_F$.
    (b)~Same information as in panel (a), taken for the constant displacement field cut $E_z = 29\,\mathrm{meV}$ as indicated by the grey arrows in \cref{fig:frg}~(a). The grey solid line indicates the DOS as the density $n/n_0$ is varied across the van-Hove line. 
    For low scales, the minimal value of $\alpha_c$ (grey dashed line) is encountered on the van-Hove line (green arrow) and the leading momentum transfer $\bvec Q_c$ is displaced from $K^{\nu}$. At larger scales, $\alpha^{\text{min}}_c$ is shifted to larger hole fillings and $\bvec Q_c$ locks back to $K^{\nu}$.
    }
    \label{fig:asymmetry}
\end{figure}

Having established the FRG phase diagram of $\theta=5.08^{\circ}$ \tWSe{} and the interplay of orbital-selective IVC-AFM order and mixed-parity $d/p$-wave superconductivity along the van-Hove line, we proceed to provide a in-depth analysis of quasi-particle properties in the ordered IVC-AFM phase by analyzing the momentum structure of the inter-valley particle-hole susceptibility as well as performing an extensive FRG+MF analysis. 
We unravel significant electronic screening contributions accumulated by particle-particle scattering within FRG that lead to a substantial suppression of effective Hubbard-like interactions. 
We further demonstrate that our microscopic, first-principle treatment captures key aspects of the experimental phase diagram including an asymmetric occurrence of IVC-AFM domains with respect to the tunable van-Hove line and the transition from a single-peak to a double-peak structure in the density of states (DOS).

\subsection{Asymmetry of particle-hole instabilities by Fermi surface broadening induced nesting} \label{sec:asymmetry}
An interesting observation of recent experiments~\cite{xia2024unconventional,guo2024superconductivity} concerns the apparent asymmetry of reconstructed metallic (insulating) regimes with respect to the displacement field tunable VHS that also manifests in the FRG phase diagram of $\theta=5.08^{\circ}$, i.e. (i) the maximal transition temperature of the IVC-AFM phase is encountered for values of the displacement field below the position of the HOVHS and (ii) instabilities are enhanced for $n < n_\mathrm{VHS}$, but diminish for $n > n_\mathrm{VHS}$.
Here, we demonstrate that in the weak-to-moderate interacting regime, this asymmetry is related to scale-dependent nesting properties of the Fermi surface~\cite{mai2022intertwined,fedor2022two} that lead to commensurability locking of particle-hole instabilities away from the van-Hove line and manifests in the inter-valley particle-hole susceptibility $\chi^{\text{FRG}}_{\uparrow \downarrow \uparrow \downarrow}(\bvec q)$.
To analyze the aforementioned asymmetric behavior, we concentrate on a single-channel RPA resummation Eq.~\eqref{eq:rpa} in the crossed particle-hole channel as main source of IVC-AFM order and calculate the critical (relative) interaction strength $\alpha_c$ that is defined by the divergence of the RPA renormalized interaction $C^{\text{RPA}}(\bvec q)$ [c.f. Eq.~\eqref{eq:rpa}], i.e. $\det[\matrixOne - \alpha_c\hat \Gamma^{0} \hat\chi^{0,C}(\bvec q)] = 0$. Here, $\Gamma^0$ is the bare interaction vertex that contains the Hubbard interaction terms defined in Eq.~\eqref{eq:fullham}.
\Cref{fig:asymmetry}~(a) displays the critical value $\alpha_c$ and the leading momentum transfer $\bvec Q_C$ of the (incommensurate) IVC-AFM state along the van-Hove line for different RG scales $\Lambda$. The latter enters the RPA theory as temperature in the Fermi-Dirac statistics and hence determines the smearing of states in the vicinity of the Fermi energy, see SM~\cite{SM} for technical details. At low scale $\Lambda = 0.1\,\mathrm{meV}$, the minimal critical interaction $\alpha_c^{\text{min}}$ (gray dashed line) is found at the HOVHS (blue circle), where the leading momentum transfer of the associated IVC-AFM state locks to $\bvec Q_C = K^{\nu}$.
%, see colorbar for the position of $\bvec Q_C$ along the high-symmetry path from $K-M$. 
This finding is consistent with the enhanced DOS at the HOVHS and the fact that the vector $\bvec q = K^{\nu}$ perfectly connects Fermi surface sheets (FS) of opposite spin-valley polarization. To this end, we show the $\Lambda$-broadened FS, i.e., all eigenenergies that reside within the energy shell $[E_F - \Lambda, E_F + \Lambda]$ around the Fermi energy $E_F$, in the lower panel of \cref{fig:asymmetry}~(a). Lowering the displacement field continuously displaces $\bvec Q_C$ towards the $M$-point and increases the critical interaction strength $\alpha_c$ as scattering between the triangular FS pockets around $K^{\nu}$ is suppressed due to the relative orientation mismatch of the van-Hove points in different spin-valley sectors.
At larger scales $\Lambda = \{0.5, 1.0 \}\,\mathrm{meV}$, however, the suppression of particle-hole instabilities for $E_z<E_z^c$ and the associated shift of $\bvec Q_C$ towards the $M$-point are counteracted by the induced smearing of the triangular FS pockets formed by the three van-Hove points. Lowering the displacement field along the van-Hove line (left panel) increases the size of the FS pockets around $K^{\nu}$ such that scattering with vector $\bvec Q_C =K^{\nu}$ between FS sheets of opposite spin-valley polarization is (re)enhanced at larger scales $\Lambda$.
As a consequence of the higher scale $\Lambda$, the minimal critical value $\alpha_c^\mathrm{min}$ moves to lower values of the displacement field along the van-Hove line. This causes an asymmetry of interaction-induced particle-hole instabilities with respect to the HOVHS.

\begin{figure*}[t]
    \centering
    \includegraphics[width=0.95\textwidth]{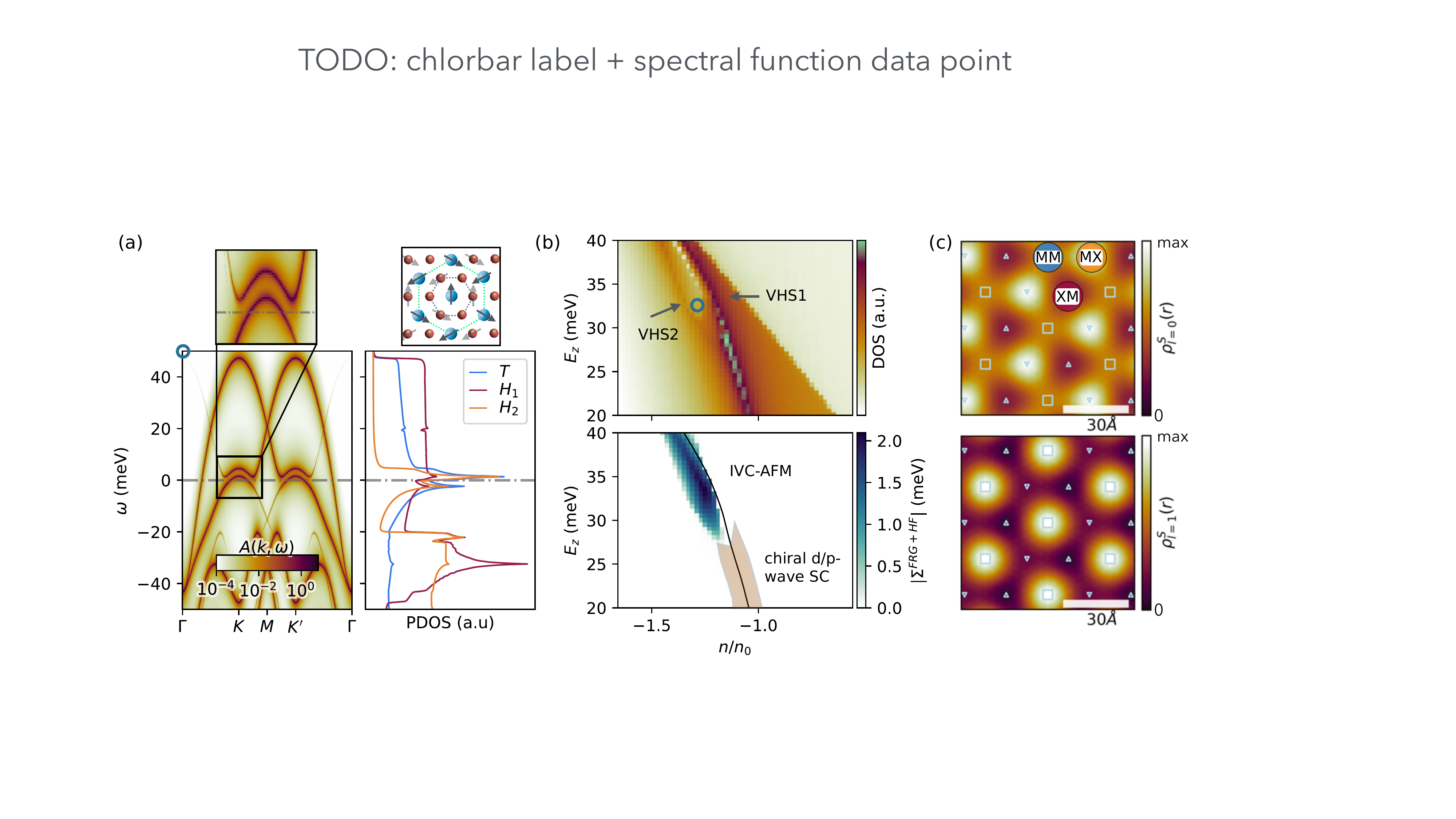}
    \caption{%
    Spectral properties of the IVC-AFM domains and VHS splitting in $\theta = 5.08^{\circ}$ \tWSe{} obtained by FRG+MF.
    (a) Momentum-resolved spectral function $A(\bvec k, \omega)$ and partial DOS (PDOS) in the IVC-AFM phase along the irreducible path of the original BZ at $E_z = 32$ meV and $n/n_0 \sim -1.27$ (blue circle).  
    The translational symmetry breaking of the IVC-AFM phase causes an effective $\sqrt{3} \times \sqrt{3}$ moiré supercell (green hexagon in inset on the upper right). 
    The IVC-AFM order parameter causes the emergence of (partial) mini-gaps around $K^{\nu}$, leading to an asymmetric splitting of the original van-Hove peak and a drastic reduction of the hole carrier density at the Fermi level (grey dashed line) that becomes visible in the PDOS. 
    (b) DOS and magnitude of the IVC-AFM order parameter in the $(E_z, n)$ phase diagram obtained for FRG+MF simulations at each point in phase space. The presence of IVC-AFM order leads to the emergence of a second van-Hove line (VHS2) associated to the hole-doped IVC-AFM state.
    The lower panel shows the amplitude of the IVC-AFM order parameter and the nearby SC regions predicted by the FRG.
    (c) Spatially-resolved spin spectral density $\rho^S_{l=0,1}(\bvec r)$ in the lower (upper) layer of \tWSe{}.
    }
    \label{fig:ivc_spectral}
\end{figure*}

A similar asymmetry is encountered when varying the hole density $n/n_0$ for constant values of the displacement field $E_z<E_z^c$ as demonstrated in \cref{fig:asymmetry}~(b). To show this, we focus on $E_z = 29\,\mathrm{meV}$ (gray dash), where the asymmetry is maximal in the FRG phase diagram as demonstrated in \cref{fig:frg}~(a). In this regime, the Fermi surface at the VHS ($n_\mathrm{VHS} \sim -1.3$, green arrow) consists of three van-Hove points that are displaced from $K^{\nu}$ such that the leading momentum transfer $\bvec Q_C$ is shifted towards the $M$-point. 
Decreasing the hole density lifts the van-Hove points and increases the size of the (open) FS pockets, similar to the observed phenomenology when varying the displacement field along the van-Hove line. For larger scales (temperatures), the broadening of the FS pockets (re)enhances scattering processes with momentum transfer $\bvec Q_C = K^{\nu}$ and the minimal critical interaction strength $\alpha_c^{\text{min}}$ is moved to hole densities $n<n_\mathrm{VHS}$. 

In conclusion, we argue that the minimal critical interaction $\alpha_c^{\text{min}}$ is shifted to lower displacement fields $E_z<E_z^c$ and lower densities $n<n_\mathrm{VHS}$ with respect to the position of the HOVHS and the displacement field tunable van-Hove line when particle-hole instabilities are encountered at non-zero temperature, or higher scales $\Lambda$. 
As the critical scale $\Lambda_c$ of particle-hole instabilities encountered in the FRG flow depends primarily on the strength of the Hubbard interactions $U_X$ in \cref{eq:fullham},
it is an indicator for how strongly coupled the system under scrutiny is. Therefore, it is expected that the observed asymmetry is enhanced when increasing the strength of the interactions relative to the bandwidth. This can be achieved by tuning the twist angle (see \cref{sec:twist}) or altering the dielectric environment, i.e., changing the distance to metallic gates. We further note that the scale-dependent nesting properties of \tWSe{} are tied to microscopic details of the fermiology of the system, particularly the formation of the triangular FS pockets around $K^{\nu}$ and the flat regions in momentum space originating from the HOVHS. This asymmetric occurrence of particle-hole instabilities and the associated commensurability locking may hence not be unique to \tWSe{}, but appear generically in systems with tunable FS pockets and HOVHS.

\subsection{FRG+MF analysis of IVC-AFM order in \texorpdfstring{\tWSe{}}{WSe2}}\label{sec:frg-mf}

To access quasiparticle properties in the ordered IVC-AFM phase for scales $\Lambda_{\text{MF}} \lesssim \Lambda_{\mathrm{c}}$ predicted by FRG, we construct a suitable bare interaction $V^{\text{MF}}$ for mean-field (MF) treatment by solving an inverse Bethe-Salpeter equation Eq.~\eqref{eq:inv-bse} in the crossed particle-hole channel ($C$) channel as outlined in Section~\ref{sec:frg}. To this end, we first investigate the orbital and momentum dependence of the effective mean-field vertex $[V^{\text{MF}}]_{\kappa \kappa'}(\bvec q)$, where again $\kappa = (\bvec k_\gamma, X, \nu_1, \nu_2)$ is a multi-index comprising the fermionic momentum variable $\bvec k_\gamma$ and Wannier function index $X$ and valley $\nu_{1,2}$. We demonstrate in the Supplementary Material~\cite{SM} that the effective mean-field vertex reconstructed from the full FRG flow can effectively be mapped to a local Hubbard-type interaction tensor of the same parametric form as Eq.~\eqref{eq:fullham} 
\begin{equation}
V^{\text{MF}} = \sum_{\nu,\bvec R}\,\sum_{X\in\{T, H_1,H_2\}} \frac{\tilde U_X}2 n_{\bvec RX}^\nu n_{\bvec RX}^{\bar\nu} \,.
    \label{eq:v-mf-U}
\end{equation}
However, opposite to the bare Wannier Hamiltonian defined in Eq.~\eqref{eq:fullham} the values of $\tilde U_X$ are substantially suppressed compared to the bare values, which we attribute to electronic screening mediated by particle-particle fluctuations included in the full FRG flow, see Eq.~\eqref{eq-frg}. We show in the Supplementary Material~\cite{SM} that the values of the effective on-site interactions $\tilde U_T = 46.7$ meV and $\tilde U_H = 60.4$ meV are barely changed as the system is tuned along the van-Hove line. We therefore adapt the above values for the remainder of this Section.

We note that it has indeed been observed in related theoretical works~\cite{munoz2025twist} that in order to reproduce the experimental phase diagram of \tWSe{} even qualitatively, the value of the effective dielectric constant must be chosen in the range of $\epsilon = 48 \ldots 60$. The FRG+MF analysis presented in this Section predicts an additional dielectric constant of $\epsilon_P \sim 2$ arising from particle-particle scattering included in the full FRG flow alone. The full-fledged FRG+MF treatment hence give insights on the discrepancy between Hartree-Fock and FRG and stress the importance of accounting for additional diagrammatic contributions beyond Hartree-Fock for quantitative comparison with experimental results in \tWSe{} and related twisted TMDs.

\subsection{Quasiparticle renormalization in the IVC-AFM phase}\label{sec:ivc}

Based on the effective mean-field vertex $V^{\text{MF}}$ defined in Eq.~\eqref{eq:v-mf-U} that is reconstructed from the full FRG flow, we investigate interaction-induced band renormalizations and spectral properties of \tWSe{} inside the ordered IVC-AFM phase.
Motivated by the Fermi surface broadening-induced nesting that pins the dominant IVC-AFM ordering to the commensurate transfer momentum $\bvec Q_c = K^{\nu}$ (c.f. discussion in Section~\ref{sec:asymmetry}), we perform a self-consistent mean-field (Hartree-Fock) decoupling of $V^{\text{MF}}$ for $\sqrt{3} \times \sqrt{3}$ moiré supercells that account for the translational symmetry-breaking of the IVC-AFM state.
The results of our FRG+MF study are summarized in Fig.~\ref{fig:ivc_spectral}. 
To access spectral properties within the ordered IVC-AFM phase, we first compute the momentum-resolved spectral function $A(\bvec k, \omega)$ and partial DOS (PDOS) in the IVC-AFM phase along the irreducible path of the original BZ at $E_z = 32$ meV and $n/n_0 \sim -1.27$, see Fig.~\ref{fig:ivc_spectral}~(a). 
The IVC-AFM order parameter causes the opening of (partial) mini-gaps around $K^{\nu}$, effectively lifting the triangular Fermi pockets that feature the highest (momentum-resolved) DOS as shown in the left panel of Fig.~\ref{fig:ivc_spectral}~(a). 
This leads to an asymmetric splitting of the original van-Hove peak and a drastic reduction of the hole carrier density at the Fermi level (grey dashed line) that becomes visible in the PDOS. 
As the IVC-AFM order parameter only has weight on the $T/H_2$-orbital, the splitting is largest in the vicinity of $K^{\nu}$, where the Fermi pockets in the normal-state are exclusively $T/H_2$-polarized, see Fig.~\ref{fig:wannier}~(c)
Meanwhile, the dispersive Fermi arcs that connect the shallow Fermi pockets around $K^{\nu}$, i.e., irreducible path from $\Gamma$ to $K^{\nu}$ remain gapless at incommensurate densities $n/n_0 \neq -1$.
Therefore, the system remains metallic despite the opening of the mini-gaps around $K^{\nu}$. 
%

%Spectral properties and depletion of spectral weight at the Fermi level
The orbital-selectivity of the IVC-AFM order parameter (only Wannier orbitals $T/H_2$ contribute) leads to a spatially inhomogeneous spin spectral density $\rho^S(\bvec r)$ defined as 
\begin{equation}
    \rho^S_l(\bvec r) = \sum_{X,  \mu \nu} \mathcal W^{X*}(\bvec r,l) \Delta^{\text{AFM}}_{X, \mu \nu} (\bvec r) \mathcal W^{X}(\bvec r,l) \,,
    \label{eq:spin-spectral-density}
\end{equation}
where $\Delta^{\text{AFM}}_{X,\mu \nu}$ is the IVC-AFM order parameter defined in Eq.~\eqref{eq:sdw_order_parameter} and $\mathcal W^X(\bvec r,l)$ is the Wannier function associated to orbital $X \in \{T, H_1, H_2\}$ in layer $l$.
We plot the layer-resolved spin spectral density in the IVC-AFM phase throughout the moiré unit cell in Fig.~\ref{fig:ivc_spectral}~(c).
Due to the $T/H_2$ polarization of the IVC-AFM order parameter, the spin spectral density is peaked in the MM/MX stacking regions in the upper layer and the MM regions in the lower layer.
This orbital selection is a direct consequence of the topological properties of the flat bands of $\theta=5.08^{\circ}$ \tWSe{} that require a multi-orbital Wannier description and should be detectable in spectroscopic measurements.

\subsection{Double-peak structure in the DOS from hole-doping the IVC-AFM state}

%VHS splitting
The Fermi surface reconstruction within the IVC-AFM phase further leads to a distinct splitting of the van-Hove line in the $E_z-n/n_0$ phase diagram as shown in Fig.~\ref{fig:ivc_spectral}~(b).
To this end, we perform a self-consistent mean-field decoupling of $V^{\text{MF}}$ for every point in the phase diagram. 
Approaching the IVC-AFM domain in the phase diagram from below, we observe that the single-peak structure of the DOS associated to the position of the VHS in the normal-state changes to a two-peak structure (VHS1/VHS2) at larger displacement field where IVC-AFM order manifests.
To understand the density dependence of the van-Hove splitting, we extract the magnitude of the IVC-AFM order parameter, i.e. the maximal value of the Hartree-Fock self-energy $|\Sigma^{\text{FRG+HF}}|$ and the position of the (reconstructed) VHS1 in the lower panel of Fig.~\ref{fig:ivc_spectral}~(b).
Inbetween VHS1 and VHS2, where the IVC-AFM has maximal amplitude, the DOS of quasiparticles is strongly suppressed due to the reduction of the hole carrier density at the Fermi level and the opening of mini-gaps around $K^{\nu}$ as demonstrated in Fig.~\ref{fig:ivc_spectral}~(a).
The region of suppressed DOS is flanked by the two reconstructed van-Hove lines and we observe that the DOS associated to the van-Hove peak of VHS2 is always weaker than the DOS of VHS1.
This is because VHS1 is in fact the extension of the VHS line already present in the normal-state, whereas VHS2 is associated to the \textit{hole-doped} IVC-AFM state and therefore only one of the asymmetrically-split van-Hove singularities that emerge within the IVC-AFM phase contribute to the DOS, c.f.~PDOS in Fig.~\ref{fig:ivc_spectral}. 
Indeed, tracing the amplitude of the IVC-AFM order parameter in the $E_z-n/n_0$ phase diagram with respect to VHS1 (black line in the lower panel of Fig.~\ref{fig:ivc_spectral}~(b)), we find that IVC-AFM order only exists to the left of VHS1. 
Meanwhile, VHS2 occurs within the ordered IVC-AFM phase and is hence a signature of the hole-doped IVC-AFM state~\cite{xia2025simulatinghightemperaturesuperconductivitymoire}.
Increasing the displacement field successively suppresses IVC-AFM order [c.f.~FRG phase diagram in Fig.~\ref{fig:frg}~(a)] such that the two VHS peaks come closer in hole density and eventually merge at large displacement fields.

%Comparison with experiment
We stress that the phase diagram obtained by the FRG+MF analysis captures the key aspects of the experimental phase diagram of \tWSe{} at $\theta=5.08^{\circ}$ outlined in Ref.~\cite{guo2024superconductivity}.
The transport measurements conducted in Ref.~\cite{guo2024superconductivity} measure the longitudinal $R_{xx}$ and Hall resistance $R_{xy}$ in the $E_z-n/n_0$ phase diagram and can pinpoint the position of the van-Hove line (increase in longitudinal resistance) as well as the `layer-polarized'{} regime at which only the $T/H_1$-orbital contribute to transport in the upper moiré band. 
At $n/n_0 < -1.1$, the experiment finds a correlated metallic state with increased longitudinal resistance that develops a two-peak structure, whereas for $n/n_0 > -1.1$ superconductivity prevails along the van-Hove line. 
Our theoretical investigation confirms the interpretation that the increased longitudinal resistivity observed experimentally can be traced back to the depletion of spectral weight at the Fermi level within the ordered IVC-AFM state.
The emergence of the characteristic second peak in the longitudinal resistivity for $n/n_0 < -1.1$ is then related to hole-doping the IVC-AFM state, where the Fermi level eventually crosses the van-Hove peak associated to the partially gapped IVC-AFM order.
This explains the asymmetry in longitudinal resistance between VHS1/VHS2.
The reconstructed IVC-AFM metal has its largest amplitude at $n/n_0 \sim -1.1$ and decreases continuously for larger values of the displacement field. 
Below the transition to the IVC-AFM state, the FRG  predicts chiral $d/p$-wave SC order as schematically summarized in the lower panel of Fig.~\ref{fig:ivc_spectral}~(b).

\begin{figure}
    \centering
    \includegraphics[width=0.9\columnwidth]{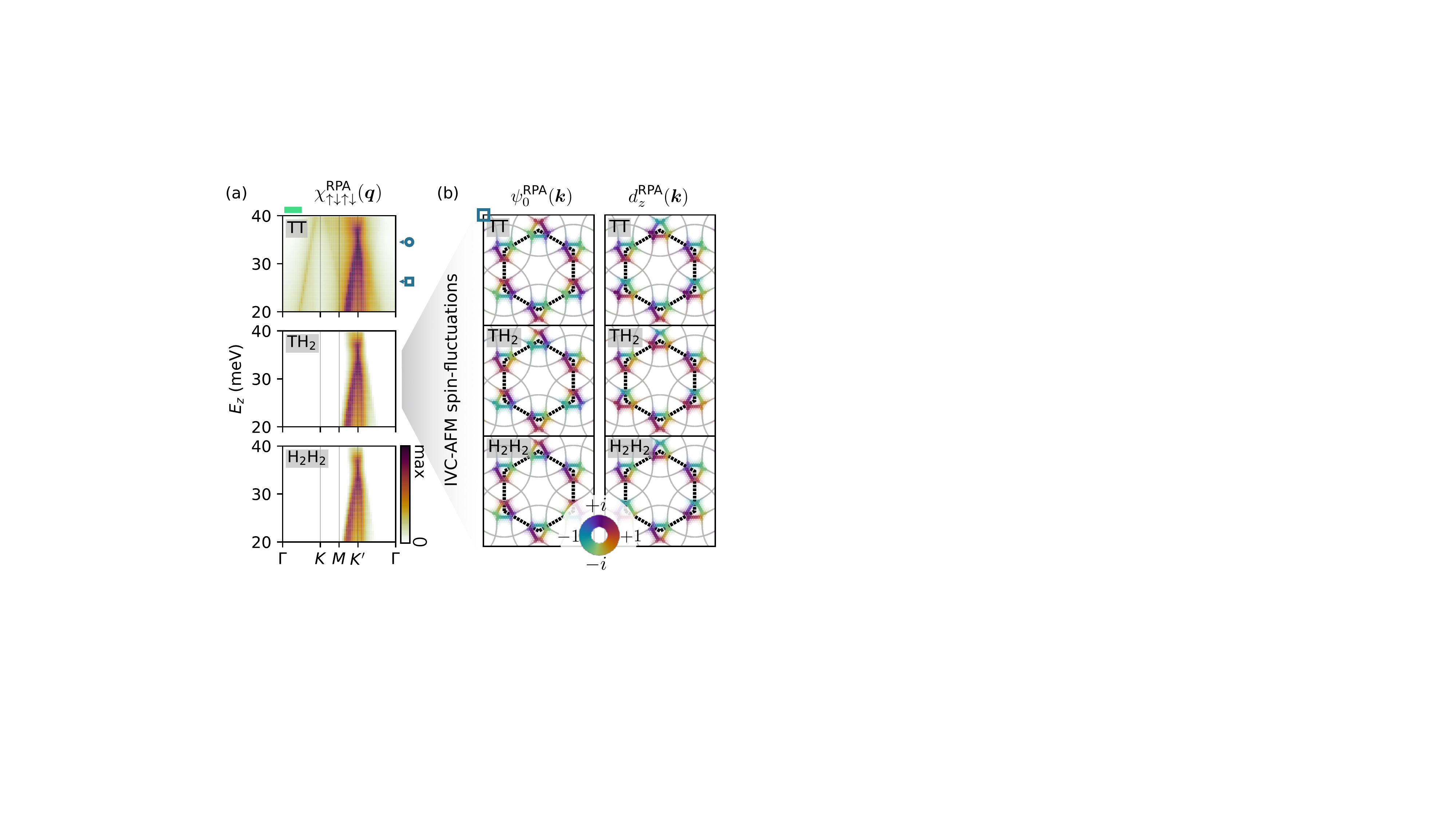}
    \caption{%
    %\textbf{
    IVC-AFM spin fluctuation mechanism to unconventional superconductivity in \tWSe{} by commensurate to incommensurate transition.
    %}
    (a)~Absolute value of the RPA-renormalized susceptibility $\chi^{\text{RPA}}_{\uparrow \downarrow \uparrow \downarrow}(\bvec q)$ at the critical interaction strength $\alpha_c U$ involving the dominant Wannier orbitals $T/H_2$. The leading momentum transfer $\bvec Q_C$ locks to $K'$ at the HOVHS (blue circle) and successively shifts towards incommensurate momenta along the high symmetry path $K'$-$M$ when lowering the external displacement field along the van-Hove line (blue square). 
    (b)~Fluctuations around the (incommensurate) IVC-AFM order give rise to chiral $d/p$-wave superconductivity comprising an equal mixture of spin-singlet $\psi_0^{\text{RPA}}(\bvec k)$ and spin-triplet $d_z^{\text{RPA}}(\bvec k)$ weight in the order parameter. For $E_z \sim 25\,\mathrm{meV}$ (blue square), the phase of the resulting order parameter is indicated by the circular colorbar and the amplitude (projected on the Fermi surface) is encoded by the opacity. 
    The superconducting order parameter only has notable weight on the triangular pockets around $K^{\nu}$, whereas the superconducting amplitude vanishes on the larger Fermi arcs connecting neighboring $K^{\nu}$ points.
    }
    \label{fig:rpa}
\end{figure}

\section{Superconductivity: Mechanism and orbital-selectivity}\label{sec:sc}

Having established a close agreement between the IVC-AFM order parameter predicted by the FRG+MF analysis and the experimental indication of $\theta=5.08^{\circ}$ \tWSe{}, we continue to analyze the superconducting order parameter predicted by the FRG analysis.
To this end, we first analyze the mechanism that drives SC order along the van-Hove line and then discuss its topological and spectral properties.

\subsection{Spin-fluctuation exchange as pairing mechanism in \texorpdfstring{\tWSe{}}{WSe2}}\label{sec:spin-fluctuations}

To unravel the microscopic mechanism that supports superconductivity along the van-Hove line, we switch off inter-channel coupling of the full FRG flow equations and thereby arrive at a single-channel RPA resummation of the interaction in the crossed particle-hole channel, c.f. Eq.~\eqref{eq:rpa}
\begin{equation}
    \label{eq:rpa-vertex}
    C^\mathrm{RPA}(\bvec q) =  \Gamma^0 \hat \chi^\mathrm{RPA}(\bvec q)  \Gamma^0 \,.
\end{equation}
This procedure segregates spin fluctuations from other possible fluctuations captured by the FRG. The resulting RPA-renormalized particle-hole susceptibility $\hat \chi^{\text{RPA}}_{\uparrow \downarrow \uparrow \downarrow}(\bvec q)$ at scale $\Lambda=0.1\,\mathrm{meV}$ is shown in \cref{fig:rpa}~(a). 
As expected from the Fermi surface sheets in the different spin-valley sectors, c.f. \cref{fig:wannier}~(c), the dominant ordering vector $\bvec Q_C$ locks to $K^{\nu}$ at the HOVHS (blue circle) and successively shifts towards incommensurate momenta on the $K^{\nu}$-$M$ line when lowering the external displacement field $E_z$. 
In this regime, the Fermi surface consists of triangular pockets around $K^{\nu}$ featuring three inequivalent van-Hove points associated to the three neighboring $M$-points, see Fig.~\ref{fig:wannier}~(c). 
Scattering between the van-Hove points enhances the particle-hole susceptibility leading to a pronounced commensurate-to-incommensurate transition of the IVC-AFM state along the van-Hove line, which is in good agreement with the interacting particle-hole susceptibility obtained from the full-fledged FRG treatment shown in \cref{fig:frg}~(b). 
To address pairing instabilities originating from fluctuations around the (incommensurate) IVC-AFM order that are present in the paramagnetic normal-state, we construct an effective pairing vertex $\Gamma^{P}(\bvec k_P, \bvec k'_P)$ from the RPA renormalized interaction in the exchange channel~\cite{ueda2004,fischer2021spin} and analyze the symmetry of the leading superconducting order parameter by solving a linearized gap equation in orbital space, see SM~\cite{SM} for technical details. As shown in \cref{fig:rpa}~(b), we find a superconducting state with mixed-parity that transforms in the two-dimensional $E$ irreducible representation ($d/p$-wave) in leading order for all values of the displacement field, exemplified for $E_z = 25\,\mathrm{meV}$ (blue square). 
The emergence of superconductivity from spin-fluctuation exchange can be understood by the real-space structure of the effective RPA pairing vertex $\Gamma^{P}(\bvec k_P, \bvec k'_P)$ that features attractive components on bonds connecting neighboring $(T, H_2)$ orbitals such to overcome the initially repulsive Coulomb interaction.
As the phenomenology of the IVC-AFM state and the symmetry of the superconducting order parameter accurately agrees with the FRG results, we argue that spin fluctuations represent the pairing glue for electron-mediated superconductivity observed in \tWSe{} along the van-Hove line.

\subsection{Orbital-selective superconductivity}\label{sec:sc_orb}
\begin{figure}
    \centering
    \includegraphics[width=0.85\columnwidth]{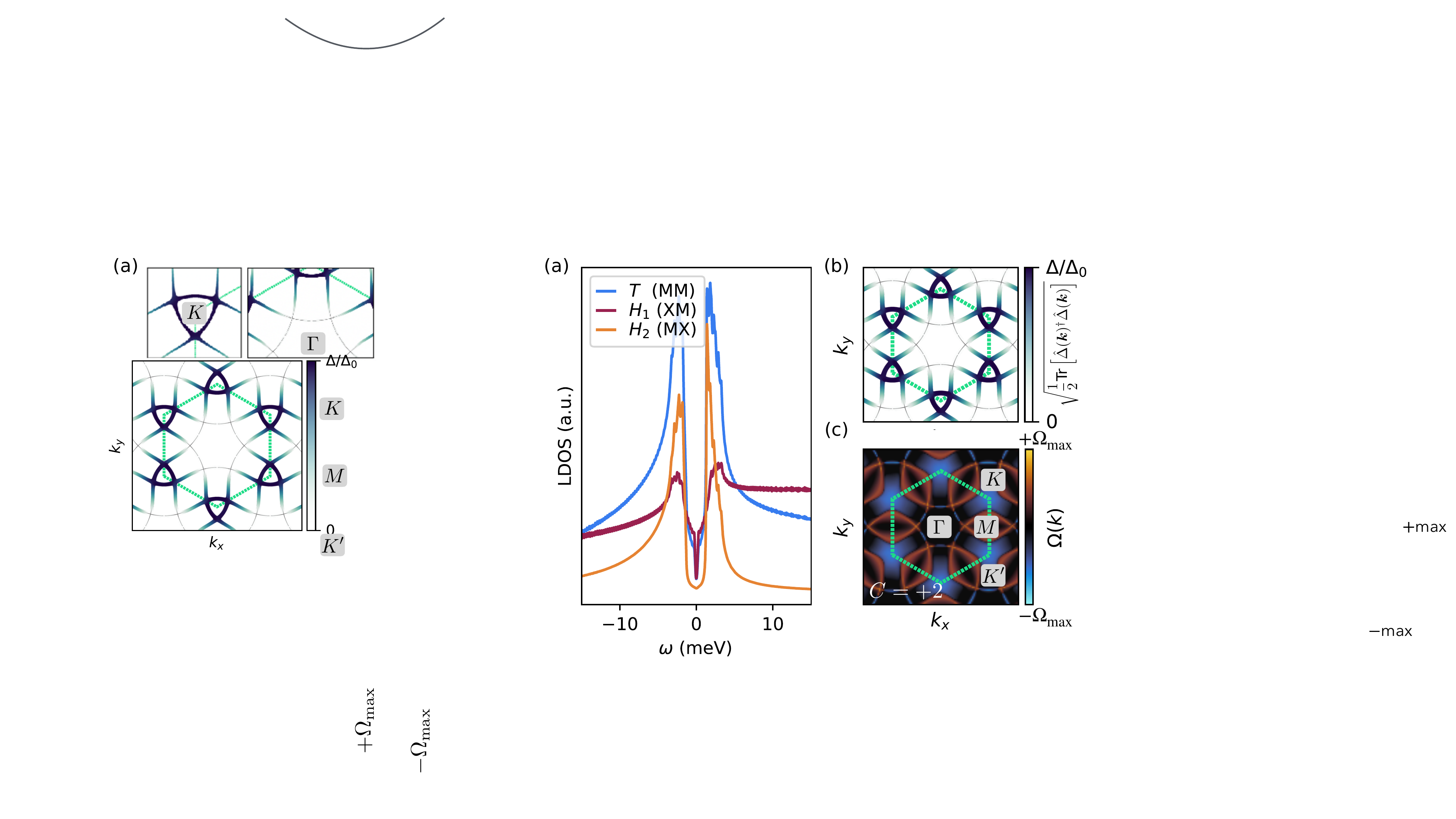}
    \caption{%
    Spectral and topological properties of the chiral $d/p$-wave SC order parameter.
    (a) Local density of states (LDOS) in the superconducting phase resolved for the Wannier orbitals $X \in \{T, H_1, H_2 \}$ that reside at the MM, XM and MX stacking regions of the moiré unit cell respectively. 
    (b) Momentum-resolved amplitde of the superconducting order parameter in the moiré Brillouin zone.
    (c) Non-abelian Berry curvature $\Omega (\bvec k)$ over the occupied Nambu bands of the Bogoliubov de-Gennes Hamiltonian. 
    The chiral $d/p$-wave SC order parameter features a non-vanishing Chern number $C=2$.
    }
    \label{fig:sc-spectral}
\end{figure}

Next, we discuss the spectral and topological properties within the ordered SC phase as summarized in Fig.~\ref{fig:sc-spectral}.
The amplitude of the superconducting order parameter defined as
\begin{equation}
|\Delta(\bvec k)|= \sqrt{\frac 12 \text{Tr} \left [ \hat \Delta^{\dagger}(\bvec k) \hat \Delta(\bvec k) \right ]} \,.
    \label{eq:sc-amplitude}
\end{equation}
is shown in Fig.~\ref{fig:sc-spectral}~(b) and is enhanced on the triangular Fermi pockets around $K^{\nu}$ that feature the largest momentum-resolved DOS and are exclusively $T/H_2$-polarized, c.f.~Fig.~\ref{fig:wannier}~(c).
As the SC order parameter only has weight on bonds including the $T/H_2$-orbital, whereas $\Delta_{H_1 X}(\bvec k) = \Delta_{X H_1}(\bvec k) = 0$ for $X \in \{T, H_1, H_2\}$, the amplitude is significantly suppressed on those Fermi surface sheets that have non-vanishing $H_1$-polarization. 
This manifests in a vanishing SC amplitude on the dispersive Fermi surface sheets that connect neighboring $K^{\nu}$ points as shown in Fig.~\ref{fig:sc-spectral}~(b).
Therefore, the orbital-selective superconducting state is effectively gapless over large parts of the Fermi surface leading to nodal signatures in spectroscopy measurements. 
To underline this statement, we calculate the local density of states (LDOS) 
in the SC phase as shown in Fig.~\ref{fig:sc-spectral}~(a).
While the $H_2$-orbital that is centered at the MX stacking regions of the moiré unit cell features a full SC gap visible by the `U'-shape of the LDOS, the MM and XM stacking regions show nodal `V'-shape characteristics. 
These nodal signatures are associated to the orbital-selectivity of the SC order parameter exhibiting strong suppression of the pairing amplitude on the dispersive Fermi arcs with mixed $T/H_1$ polarization. 
We therefore claim that the SC order parameter, albeit being fully gapped around $K^{\nu}$, features nodal signatures in spectroscopy measurements. 

%Topological properties
The chiral superconducting state of $\theta=5.08^{\circ}$ \tWSe{} belong to the A Cartan class and possesses a $\mathbb{Z}$ topological invariant as outlined in Section~\ref{sec:frg}. 
Here, we explicitly calculate the non-abelian Berry curvature 
\begin{align}
    \bvec \Omega (\bvec k) &{}= - \text{Im} \big \{ \log \det_{\mu\nu} \langle \nabla_{\bvec k} u_{\mu}(\bvec k) | \times | \nabla_{\bvec k} u_{\nu}(\bvec k) \rangle  \big \} 
    %\tilde \Omega(\bvec k) &{}= \langle \partial_{\bvec k} u_{\mu}(\bvec k) | \times | \partial_{\bvec k} u_{\nu}(\bvec k) \rangle \,,
    \label{eq:non-abelian-berry}
\end{align}
where $u_{\mu}(\bvec k)$ is the (Nambu) Bloch wave function at crystal momentum $\bvec k$ and the band index $\mu$ comprises all occupied states of the Nambu Hamiltonian. The Chern number is then obtained by $C= 1/(2\pi) \sum_{\bvec k} \Omega_z(\bvec k)$.
We show the distribution of Berry curvature within the moiré BZ in Fig.~\ref{fig:sc-spectral}~(c).
Berry curvature predominantly accumulates around $K^{\nu}$, where the SC order parameter fully gaps the system and along the dispersive Fermi arcs, along which the SC amplitude is drastically suppressed.

\section{Influence of long-ranged interactions on the low-energy phase diagram}\label{sec:long}
Besides the dominant Hubbard interaction terms acting on the triangular ($T$) and honeycomb sites ($H_{1,2}$) of the effective Wannier model defined in \cref{eq:fullham}, the ratio between the Wannier function spread $\Omega \sim |\bvec a_M| = 3.74\,\mathrm{nm}$ and the distance to the metallic gates $\xi \sim 10\,\mathrm{nm}$ on either side of the sample suggests that longer-ranged density-density interactions impact the low-energy physics of $\theta = 5.08^{\circ}$ \tWSe{}. Meanwhile, exchange and pair-hopping contributions between neighboring Wannier orbitals are negligible as expected from the exponential localization of the Wannier orbitals, see SM~\cite{SM}. 
Therefore, we next complement the Hubbard Hamiltonian \cref{eq:fullham} by density-density terms that are obtained when projecting the dual-gated Coulomb interaction to the Wannier basis:
\begin{equation}
H^V= H + \sum_{\nu,\mu} \!\!\! \sum_{\substack{X,X'\in \\ \{T, H_1,H_2\} } } \!\! \sum_{\bvec R, \bvec R'}  V_{XX'}^{\bvec R \bvec R'}n^\nu_{\bvec R X} n^\mu_{\bvec R'X'} \,.
\label{eq:ham_v}
\end{equation}
In the spirit of an \emph{ab-initio} characterization of correlated phases in \tWSe{}, we consider the full real-space dependence of $V_{XX'}^{\bvec R \bvec R'}$ in this Section. The largest of these density-density terms are the nearest-neighbor honeycomb-triangular (honeycomb-honeycomb) density-density interactions $V_{TH} \approx 33\,\mathrm{meV}$ ($V_{HH} \approx 28\,\mathrm{meV}$).

\subsection{Influence of long-ranged density-density interactions on the position of the van-Hove line and the layer-polarization regime.} \label{sec:hartree}
\begin{figure}[t]
    \centering
    \includegraphics[width=0.83\columnwidth]{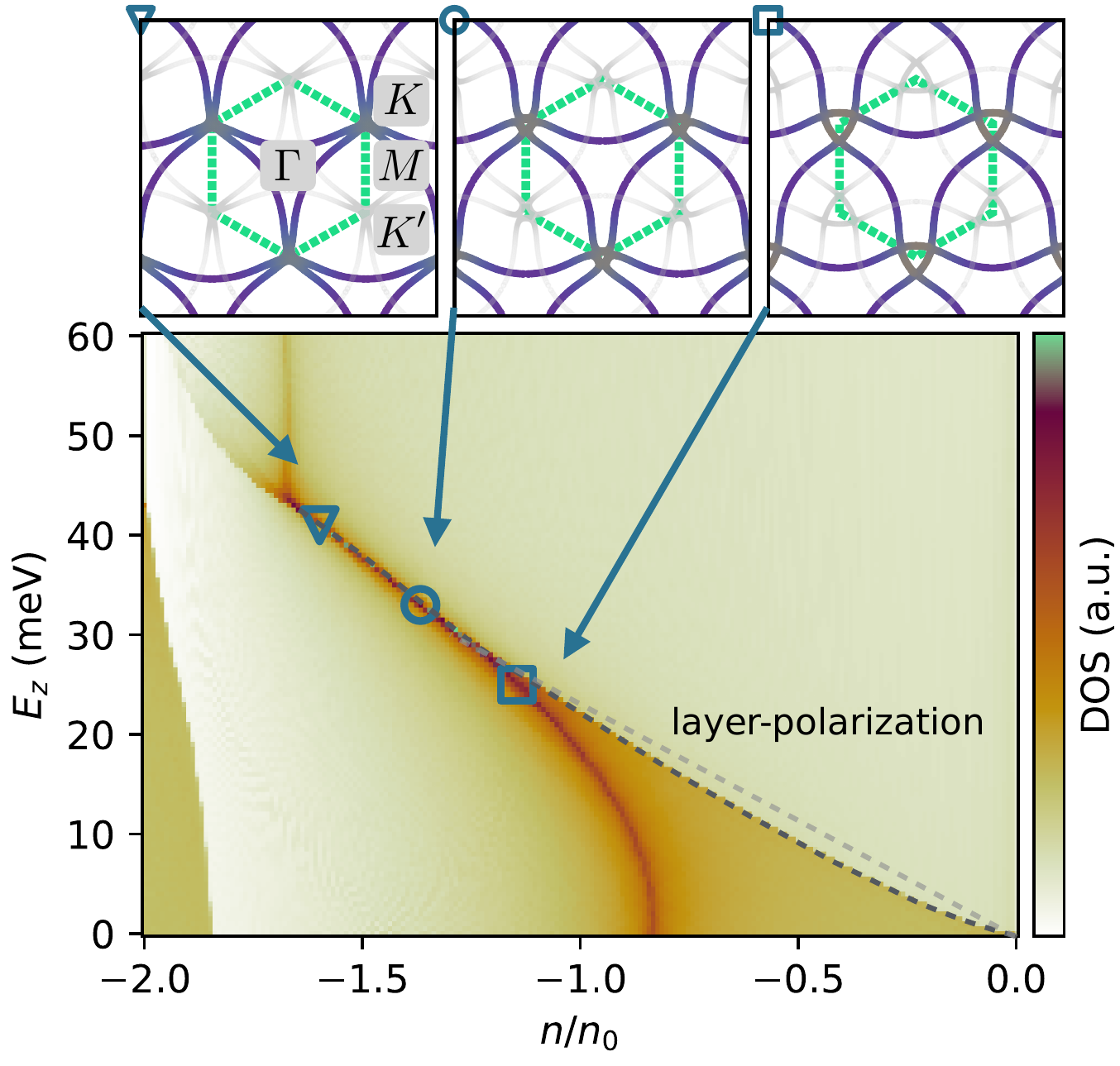}
    \caption{%
    Symmetry-unbroken band renormalizations in the presence of long-ranged density-density interactions. 
    The DOS is obtained at each point in the $(E_z,n)$ phase diagram by a FRG flow of the self-energy.
    The critical displacement field $E_Z^{\text{lp}}$ at which the system enters the 'layer-polarization' regime (black dots) and only the Wannier orbitals $T,H_1$ contribute to the Fermi surface picks up a non-linear dependence $E_z^{\text{lp}} \sim [n/n_0]^2$. 
    To see this, we plot the layer-polarization line (grey dots) in the uncorrelated normal-state from Fig.~\ref{fig:wannier}~(c) as a reference.
    The fermiology along the tunable van-Hove line (inset) resembles the uncorrelated normal-state.
    }
    \label{fig:hf-long}
\end{figure}

Before turning the discussion to the impact of long-ranged Coulomb interactions on the IVC-AFM order and superconducting regions in the phase diagram of $\theta=5.08^{\circ}$ \tWSe{}, we first investigate the effect of symmetry-unbroken moiré band renormalizations as function of external displacement field and filling. Hartree-induced band renormalizations have been shown to strongly renormalize the low-energy flat bands in twisted multilayer graphenes~\cite{Guinea2018,Rademaker2019,Cea2019,goodwin2020hartree,lewandowski2021does,fischer2022unconventional,choi2021interaction,turkel2022orderly} leading to the pinning of the van-Hove singularities of the flat moiré bands to the Fermi level such to enhance the ordering tendency of the system over extended filling regions.
Here, we analyze the impact of symmetry-unbroken Hartree-Fock induced band renormalization by starting from the Hamiltonian $H^V$ defined in Eq.~\eqref{eq:ham_v} and including the (static) self-energy $\Sigma(\bvec k)$ within the FRG formalism. Therefore, we effectively account for all Hartree-Fock type diagrams and their renormalization of the single-particle bandstructure, see Supplementary Material for details.
To disentangle the role of interaction-induced band renormalizations and the formation of electronic order, we neglect the flow of the two-particle interaction vertex $\Gamma^{(4)}$ ($\Sigma$-flow) and study the impact of symmetry-unbroken self-energy corrections on the DOS in the phase diagram of displacement field and filling as showin in Fig.~\ref{fig:hf-long}. 

First and foremost the van-Hove line is more strongly coupled to the layer-polarized regime (black dashed line), in which the system is exclusively polarized on the $T/H_1$-orbital leading to a sudden suppression of the DOS. 
This pinning of the van-Hove line with maximal DOS and the layer-polarization line is facilitated by an additional band flattening near the $K,K'$ points of the moiré BZ in the respective density regime, see Supplementary Material~\cite{SM} for plots of the renormalized bandstructure as function of external displacement field and hole filling.
Meanwhile, the fermiology along the van-Hove line is not altered by the symmetry-unbroken Hartree-Fock corrections as demonstrated in the upper panel of Fig.~\ref{fig:hf-long}~(a) that show the Fermi surface on the van-Hove line for the same displacement fields as the normal-state DOS shown in Fig.~\ref{fig:wannier}.
In the normal-state (black dashed line), the critical displacement field $E_Z^{\text{lp}}$ at which the system enters the 'layer-polarization' regime (grey dots) and only the Wannier orbitals $T,H_1$ contribute to the Fermi surface is a strictly linear function of the external displacement field as expected from the interlayer potential difference $E_z$ encoded in the multi-orbital Wannier Hamiltonian Eq.~\eqref{eq:fullham}.
In the presence of long-ranged density-density interaction, the critical displacement fields picks up a non-linear dependence $E_z^{\text{lp}} \sim [n/n_0]^2$ at small densities $-1<n/n_0<0$, in well agreement with experimental findings~\cite{guo2024superconductivity,xia2024unconventional,xia2025simulatinghightemperaturesuperconductivitymoire}.

\subsection{FRG phase diagram: Formation of spin-bond order} \label{sec:long_frg}

\begin{figure}[t]
    \centering
    \includegraphics[width=\columnwidth]{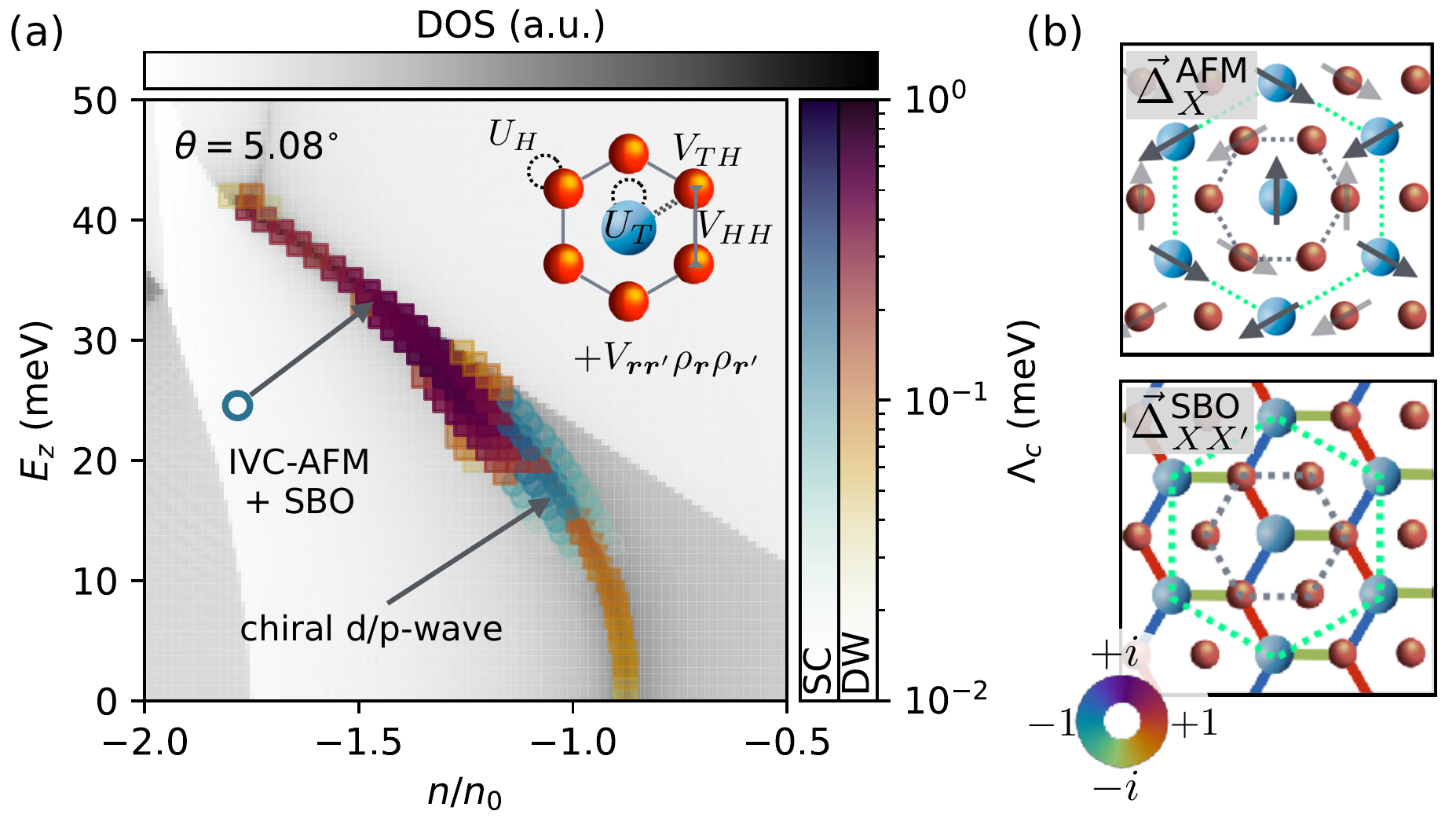}
    \caption{%
    %\textbf{
    Influence of long-ranged density-density interactions on the low-energy phase diagram of $\theta = 5.08^{\circ}$ \tWSe{}.
    %}
    (a)~FRG phase diagram as function of holes per moiré unit cell $n/n_0$ and external displacement field $E_z$. The on-site IVC-AFM order parameter mixes with a spin-bond order (SBO) parameter between the $T$/$H_2$ orbitals that transform in the $A$-IR of $C_{3z}$. This raises the critical scale $\Lambda_{\mathrm c}$ of the density wave (DW) instabilities in the phase diagram and enhances the critical temperature of the superconducting phase.
    (b)~Real-space structure of the particle-hole instabilities. The (magnetic) Wigner-Seitz cell is indicated by the (green) gray dashed line.
    }
    \label{fig:frg-long}
\end{figure}

As the symmetry-unbroken band renormalizations from long-ranged density-density interactions only yield minor quantitative corrections to the position of the DOS in the phase diagram of displacement field and filling, we proceed by studying the FRG phase diagram for $\theta = 5.08^{\circ}$ \tWSe{} as shown in \cref{fig:frg-long}~(a). While the qualitative features of the phase diagram resemble those of the pristine Hubbard model \cref{eq:fullham}, the phase boundaries and critical scales of the emergent particle-hole (particle-particle) instabilities are altered.
The observed quantitative changes arise because the inter-site interaction causes  non-local particle-hole pairs to get admixed to the local IVC-AFM order parameter. The nearest-neighbor repulsion between triangular ($T$) and honeycomb ($H_2$) orbitals supports the formation of a spin/valley-bond order (SBO)
\begin{equation}
    \vec\Delta^\mathrm{SBO}_{\bvec R,\bvec R'} \propto e^{i\bvec Q_C\cdot\bvec R} \braket{c^{\nu\dagger}_{\bvec R,T}\vec\sigma^{\nu\nu'}c^{\nu'\vdagger{}}_{\bvec R',H_2}} + \mathrm{h.c.}
    \label{eq:sbo_order_parameter}
\end{equation}
%
%The spin/valley-bond order parameter (SBO) 
The SBO transforms within the trivial $A$-irreducible representation (IR) of the little group $C_{3z}$ at $K^{\nu}$ and is therefore symmetry-allowed to mix with the IVC-AFM state $\Delta^\mathrm{AFM}$
driven by the on-site Hubbard repulsion. In \cref{fig:frg-long}~(b), we present a sketch of the magnetization patterns in the IVC-AFM+SBO phase comprising an on-site $120^{\circ}$ spin configuration as well as the bond order component between $T$-$H_2$. Note that due to the $p_{\nu}$-orbital character of the $H_2$-orbital, the bonds feature a relative phase of $\omega = e^{2 \pi i/3}$ between $C_{3z}$-related $H_2$-orbitals. 
The gain in free energy arises because the projection of the bare nearest-neighbor repulsion $V_{TH(HH)}$ to the crossed particle-hole channel lowers the energy of states with an SBO component of the order:
\begin{equation}
\Delta^{\text{SBO}*} \circ C \big [\Gamma^{(4)}_{\Lambda = \infty} \big ] \circ \Delta^{\text{SBO}} > 0 \,,
\end{equation}
see SM~\cite{SM} for details.
The non-vanishing SBO components in the order parameter hence promote the formation of particle-hole pairs at larger critical scales $\Lambda_c$ as demonstrated in \cref{fig:frg-long}~(a) [compare with \cref{fig:frg}~(a)]. We therefore argue that the particle-hole instability experiences an effective anti-screening by virtue of long-ranged density-density type interactions. The coupling to the on-site IVC-AFM order that drives the chiral $p/d$-wave pairing instability causes the maximal critical temperature to rise to $\Lambda_c \sim 1\,\mathrm{K}$. Meanwhile, the size of the SC region in the phase diagram is decreased due to the competition with the IVC-AFM state.

\section{Evolution of the IVC-AFM order and superconductivity across twist angles} \label{sec:twist}

\begin{figure*}[t]
    \centering
    \includegraphics[width=\textwidth]{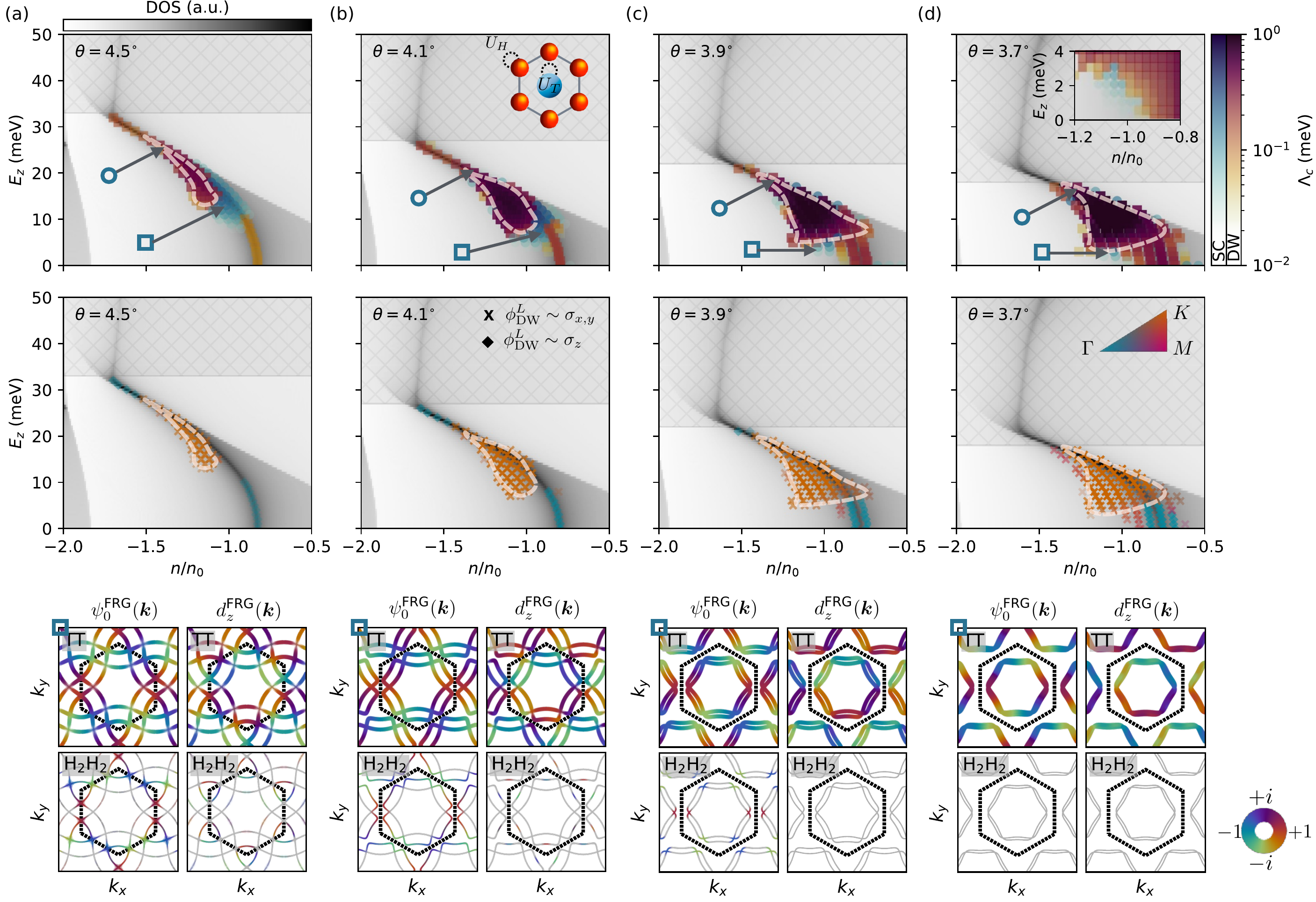}
    \caption{%
    Evolution of IVC-AFM domains and SC order in the low-energy phase diagram of \tWSe{} in the twist angle regime $\theta = 3.7^{\circ} \dots 4.5^{\circ}$.
    %}
    (a)~FRG phase diagram of $\theta = 4.5^{\circ}$ \tWSe{}. The upper row shows the critical scale $\Lambda_{\mathrm{c}} \sim T_{\mathrm{c}}$ of the leading spin/charge density wave (DW, red) and superconducting instability (SC, blue).
    The DOS of the normal-state (grey) is given as reference and the blue circle indicates the position of the HOVHS. The grey-hatched area represents the experimentally inaccessible regime.
    The middle panel shows the leading FRG eigenvector $\phi^L_{\kappa}(\bvec q_C)$ for all DW instabilities; color-coding the evolution of the leading momentum transfer $\bvec q_C$ in the irreducible wedge of the moiré Brillouin zone and differentiating between IVC [$\phi^L_{\kappa}(\bvec q_C)\sim\sigma_{x,y}$] or valley-polarized (VP) [$\phi^L_{\kappa}(\bvec q_C)\sim\sigma_{z}$] order.
    The lower panel shows the amplitude (opacity) and phase (color) of the SC order parameter on the Fermi surface (grey line) segregated into singlet (triplet) components $\psi_0(\bvec k)$ ($d_z(\bvec k)$) on the $T/H_2$-orbitals. 
    (b,c,d)~FRG phase diagrams for $\theta=4.1^{\circ}$, $\theta=3.9^{\circ}$ and $\theta=3.7^{\circ}$.
    Regions of IVC-AFM order (convex hull) evolve to smaller values of the displacement field $E_z$ and bend towards $n/n_0 \sim -1$, causing a pronounced asymmetry with respect to the van-Hove line (grey).
    Regions of chiral $d/p$-wave SC order are shrunk in size and are eventually displaced from the van-Hove line in favor of IVC-AFM order at $\theta\sim 4.0^{\circ}$. 
    In this regime, SC domains move towards $n/n_0 \sim -1$ and their critical temperature is decreased due to the change in fermiology and the suppression of DOS away from the van-Hove line.
    Chiral $d/p$-wave SC order remains the leading superconducting instability, albeit the SC amplitude is constrained to the $T$-orbital signaling a transition from multi-orbital to single-orbital SC order.
    }
    \label{fig:otherangle}
\end{figure*}

After having established our methodology for $\theta=5.08^{\circ}$ \tWSe{}, we map out the FRG phase diagrams in the twist angle regime $\theta = \{3.7^{\circ}, 3.9^{\circ}, 4.1^{\circ}, 4.5^{\circ}\}$ as summarized in \cref{fig:otherangle}~(a-d).
Decreasing the twist angle quenches the kinetic energy scales such that the moiré bands react more sensitively to the external displacement field. 
As a result, the position of the HOVHS is moved to smaller values of $E_z$, see blue circle in \cref{fig:otherangle}~(a-d).
As the ratio between the Wannier function spread and the gate screening distance $\Omega/\xi$ decreases with the twist angle, long-ranged density-density interaction between the Wannier orbitals are successively suppressed and we therefore resort to the Hubbard Hamiltonian defined in Eq.~\eqref{eq:fullham}. 
We provide a detailed overview of the twist angle dependence of the DOS and the values of the density-density interaction tensor in the SM~\cite{SM}.
At small twist angles, we only perform FRG simulations for experimentally relevant values of the displacement field corresponding (black dashed line).

Decreasing the twist angle further increases the ratio $U/W$ and places the system in a moderate-to-strongly interacting regime. 
As a consequence, the emergent IVC-AFM order manifests at larger scales $\Lambda_c$ indicating a higher critical temperature of magnetic ordering. Consistent with the analysis brought forward in \cref{sec:asymmetry}, this enhances the asymmetry of the IVC-AFM order with respect to the van-Hove line as demonstrated in \cref{fig:otherangle}~(a-d).
At $\theta=4.5^{\circ}$, the maximal critical scale $\Lambda_c$ of the IVC-AFM phase is found around $E_z \sim 18\,\mathrm{meV}$ and is clearly displaced from the van-Hove line.
At $\theta = 4.1^{\circ}$, the IVC-AFM order extend towards the commensurate hole filling of $n/n_0 \sim -1$ and replaces former superconducting regions along the van-Hove line.
At half-filling $n/n_0 \sim -1$, quasiparticle renormalizations in the IVC-AFM phase can \emph{fully} gap the Fermi surface, opposite to the scenario at larger twist angles where IVC-AFM order at incommensurate densities $n/n_0 \neq -1$ reconstructs the Fermi surface and the system is a partially-gapped antiferromagnetic metal, c.f. Fig.~\ref{fig:ivc_spectral}~(a).
The transition from a partially-gapped IVC-AFM metal to a fully-gapped insulating state in the intermediate twist angle regime $\theta \sim 4.5^{\circ}$ as predicted by our theory is in line with previous experimental measurements~\cite{xia2024unconventional,guo2024superconductivity,xia2025simulatinghightemperaturesuperconductivitymoire}.

The asymmetric bending of IVC-AFM domains is further enhanced for the smaller twist angles $\theta=3.9^{\circ}$ and $\theta=3.7^{\circ}$ such that SC order is eventually completely displaced from the van-Hove line as shown in Fig.~\ref{fig:otherangle}~(c,d).
At small twist angles and small values of the displacement field $E_z < 8$ meV, additional magnetic phases emerge in the vicinity of the van-Hove line. 
To analyze these states, we provide a full characterization of the leading FRG eigenvector $\phi^L_{\kappa}(\bvec q_C)$ in the crossed particle-hole channel ($C$) by color-coding the evolution of the leading momentum transfer $\bvec q_C$ in the irreducible wedge of the moiré Brillouin zone. We further distinguish between IVC order [$\phi^L_{\kappa}(\bvec q_C)\sim\sigma_{x,y}$] or valley-polarized (VP) [$\phi^L_{\kappa}(\bvec q_C)\sim\sigma_{z}$] solutions as shown in the middle row of Fig.~\ref{fig:otherangle}.
Along the van-Hove line and for small values of the displacement field, $\theta = 3.7^{\circ}$ \tWSe{} features domains of ferromagnetic VP order ($\bvec q_C = \Gamma$, blue), albeit their critical temperature is lower compared to the IVC-AFM domains.
In their vicinity, smaller domains of striped-IVC order ($\bvec q_C = M$, red) emerge. For larger values of the displacement field, these domains smoothly transition into regions of incommensurate IVC-AFM order ($\bvec q_C = K \to \Gamma$, darkorange) until eventually recovering commensurate IVC-AFM order ($\bvec q_C = K$, orange) that follows the van-Hove line.
For larger twist angles, the additional magnetic phases are weakened and only the IVC-AFM order survives in the vicinity of the van-Hove line.

\subsection{Evolution of SC order towards smaller twist angles: Multi-orbital to single-orbital SC order}

At $\theta \sim 3.9^{\circ}$, the superconducting region in the phase diagram is displaced from the van-Hove line and moves towards $n/n_0 \sim -1$ and increasingly smaller values of $E_z$.
Away from the van-Hove line, the fermiology of the system changes from the triangular Fermi pockets centered around $K^{\nu}$ to a single connected hexagon interleaved with the moiré BZ, see Fig.~\ref{fig:otherangle}.
The detuning from the van-Hove line causes a decrease in the DOS that leads to a sudden decrease of the critical scale $\Lambda_{\mathrm c}$ associated to the SC order parameter, albeit the system is tuned into a more strongly-correlated regime. 
We observe that the symmetry classification and topological properties of the superconducting state are not altered and the leading pairing instability across all twist angles remains the chiral, mixed-parity $d/p$-wave. 
However, while at $\theta =5.08^{\circ}$ the amplitude on the $T/H_2$ bonds is nearly equal, the SC order parameter at lower twist angles is only facilitated by the $T$ orbital that resides in the $MM$ stacking regions of the superlattice. 
Therefore, we claim that the superconducting order parameter changes from a multi-orbital to single-orbital character as the twist angle is decreased.
The SC state at lower twist angles may be compatible with early theoretical works based on the single-band `moiré-Hubbard model'~\cite{wu2019topological}, where superconductivity has been reported away from the van-Hove line at hole densities $n/n_0 \sim -1$~\cite{klebl2023competition}.

\section{Conclusion and outlook}

In this work, we provide a complete and unbiased characterization of correlated states in $\theta = 3.7^{\circ} \dots 5^{\circ}$ \tWSe{} from \emph{first principles} by studying the impact of gate-screened Coulomb interactions with state-of-the-art functional renormalization group techniques. Our analysis reveals that inter-valley coherent anti-ferromagnetic order (IVC-AFM) and orbital-selective, chiral $d/p$-wave superconductivity collaborate along the gate-tunable van-Hove line. The superconducting pairing glue for the unconventional SC state is provided by spin-fluctuations arising in regions of incommensurate IVC-AFM order below the HOVHS, where the Fermi surface pockets around the valleys $K^{\nu}$ are broadened and therefore weaken scattering between opposite spin-valley sectors. 
We attribute the asymmetry of particle-hole instabilities observed in experiment~\cite{xia2024unconventional, guo2024superconductivity} to the formation of hole pockets around $K^{\nu}$ that are broadened at larger scales (temperature) such that the dominant ordering vector locks back to $K^{\nu}$ at positions that are offset from the van-Hove line~\footnote{We foresee this scale-dependent nesting as a general feature of weak-to-moderately interacting fermions with HOVHS.}. As the hole pockets are equally polarized on the $T/H_2$-orbital, this stresses the importance of describing the low-energy physics of \tWSe{} in the multi-orbital setup presented in this manuscript.
By performing a comprehensive FRG+MF analysis within the ordered IVC-AFM phase, we show that our calculations can explain pivotal aspects of the experimental phase diagram including an asymmetric occurrence of IVC-AFM domains with respect to the tunable van-Hove line, the transition from a single-peak to a double-peak structure in the density of states (DOS) and the subtle interplay between IVC-AFM order and superconductivity along the van-Hove line.
Our numerical framework unveils that particle-particle screening leads to a significant suppression of the bare down-folded Coulomb interaction, which explains the overestimation of electronic order in Hartree-Fock based approaches that do not resolve vertex corrections.
We further provided an in-depth analysis of spectral properties in the SC phase and unravel a strong orbital-selectivity that manifests in nodal signatures in the local density of states (LDOS) and is related to the topological properties of the Wannier model.
Further, we demonstrate that long-ranged density-density interactions do not alter the phase diagram of \tWSe{} qualitatively, but change the phase boundaries and critical temperature due to the coupling of the IVC-AFM order parameter to a spin/valley bond order (SBO) that acts as an interaction-induced spin-orbit coupling term.
We therefore argue that the presence of longer-ranged density-density interactions will enhance the critical temperature $T_c$ of the SC phase, however, decrease its overall size in the phase diagram due to the competition with the IVC-AFM state.

Our theoretical predictions provide guidance for future experiments aiming to verify the nature of the correlated states in \tWSe{}. Besides magnetic circular dichroism (MCD) measurements~\cite{xia2025simulatinghightemperaturesuperconductivitymoire} of the insulating state at $n/n_0=1$ in 4.6$^\circ$ \tWSe{} suggestive of inter-valley coherent order, future experiments could probe the inhomogeneous spin spectral density in the IVC-AFM phase associated to the multi-orbital physics via local spectroscopy techniques like (magnetic) scanning tunneling microscope~\cite{nuckolls2023quantum}.
Another promising route is the newly developed quantum twisting microscope~\cite{inbar2023quantum,xiao2025interacting} that could be exploited to directly probe the momentum-resolved spectral function $A(\bvec k, \omega)$ in the correlated metallic (insulating) IVC-AFM phase as provided in Fig.~\ref{fig:ivc_spectral}.
As the chiral $d/p$-wave superconductor descends from the IVC-AFM phase and is of Ising-type, i.e. Cooper pairs are formed by electrons from different valleys with opposite spins due to strong intrinsic Ising-SOC, we expect a weak (strong) dependence of the SC state subject to in-plane (out-of-plane) magnetic fields~\cite{lu2015evidence, wang2019type,falson2020type} with possibly large Pauli-limit violations for in-plane magnetic fields.

Finally, we show that as the twist angle is decreased to $\theta \sim 4.0^{\circ}$ the regions of SC order are displaced to smaller values of the external displacement field ($5\,\mathrm{meV}<E_z<10\,\mathrm{meV}$) and to hole densities closer to $n/n_0\sim -1$ in good agreement with recent experimental measurements~\cite{xia2025simulatinghightemperaturesuperconductivitymoire}. 
For smaller twist angles $\theta \lesssim 4.0^{\circ}$, SC order is eventually displaced from the van-Hove line, which leads to a suppression of the critical temperature due to the reduction in DOS.
It is interesting to consider these results in light of the finding of Xia~\emph{et.~al.}~\cite{xia2024unconventional} that in a $\theta=3.65^\circ$ angle device, a superconducting region exists very near $n/n_0=1$ and vanishing displacement field, giving way to an insulating phase as the displacement field is increased.

We further note that the quantitative validity of the static four-point FRG employed in this work is bound to the weak-to-moderate interacting regime $U/W<1$ that seems relevant for the large-to-intermediate twist angle regime considered within this work. By further decreasing the twist angle, we expect that the system is eventually tuned into a strongly interacting regime, in which dynamical correlations gain in importance~\cite{ryee2023switching}. 
One may ask, for example, whether this superconductor-insulator transition observed in the $3.65^\circ$ twist angle device can be understood in terms of the weak/intermediate coupling theory presented here or has properties inconsistent with our theory, requiring consideration of strong coupling Mott/spin liquid physics. It is thus a highly anticipated avenue of future research to treat this regime with state-of-the art methods such as, e.g. DMFT, DMFT+FRG~\cite{taranto2014infinite} or slave Bosons to learn about the interplay of Mott insulators and the ordered IVC-AFM (chiral $d/p$-wave superconductor) proposed by FRG. 
The methodological framework developed here can further be applied to tackle related questions concerning the interplay of magnetic order and superconductivity in rhombohedral multilayer graphene~\cite{zhou2021superconductivity,han2024sig} via effective multi-orbital Wannier models as e.g. outlined in Ref.~\cite{fischer2024supercell} or to investigate the role of iso-spin fluctuations in twisted bilayer graphene by treating electronic correlations in multi-orbital
Wannier models~\cite{song2022magic,carr2019wannier} with DMFT~\cite{datta2023heavy,rai2023dynamical} as starting point.

\begin{acknowledgments}
This work was supported by the Excellence Initiative of the German federal and state governments, the Ministry of Innovation of North Rhine-Westphalia and the Deutsche Forschungsgemeinschaft (DFG, German Research Foundation). 
AF and DMK acknowledge funding by the DFG  within the Priority Program SPP 2244 ``2DMP'' -- 443274199.
LK acknowledges support from the DFG through Project-ID 258499086 -- SFB 1170 and through the W\"urzburg-Dresden Cluster of Excellence on Complexity and Topology in Quantum Matter -- ct.qmat, Project-ID 390858490 -- EXC 2147.
LK and TOW greatfully acknowledge support from the DFG through FOR 5249 (QUAST, Project No. 449872909) and SPP 2244 (Project No. 422707584). SR is supported by the DFG research unit FOR 5242 (WE 5342/7-1, project No. 449119662).
TOW is supported by the Cluster of Excellence ``CUI: Advanced Imaging of Matter'' of the DFG (EXC 2056, Project ID 390715994).
AR acknowledges support by the European Research Council (ERC-2015-AdG694097), the Cluster of Excellence ‘Advanced Imaging of Matter' (AIM), Grupos Consolidados (IT1249-19) and Deutsche Forschungsgemeinschaft (DFG) -- SFB-925 -- project 170620586.
LX acknowledges supported by the National Key Research and Development Program of China (Grant No. 2022YFA1403501), Guangdong Basic and Applied Basic Research Foundation (Grant No. 2022B1515120020), the National Natural Science Foundation of China (Grant No. 62341404), Hangzhou Tsientang Education Foundation and the Max Planck Partner group programme.
AJM acknowledges support from Programmable Quantum Materials, an Energy Frontier Research Center funded by the U.S. Department of Energy (DOE), Office of Science, Basic Energy Sciences (BES), under award (DE-SC0019443).
DMK and AR acknowledge support by the Max Planck-New York City Center for Nonequilibrium Quantum Phenomena. Computations were performed on the HPC system Ada at the Max Planck Computing and Data Facility. 
The Flatiron Institute is a division of the Simons Foundation.
\end{acknowledgments}

\section*{Data availability}
The data that support the findings of this article are openly available~\cite{zenodo}.

\bibliography{main}

\clearpage
%\includepdf[pages=-]{supp.pdf}

\foreach \x in {1,...,20}
{%
\clearpage
\includepdf[pages={\x}]{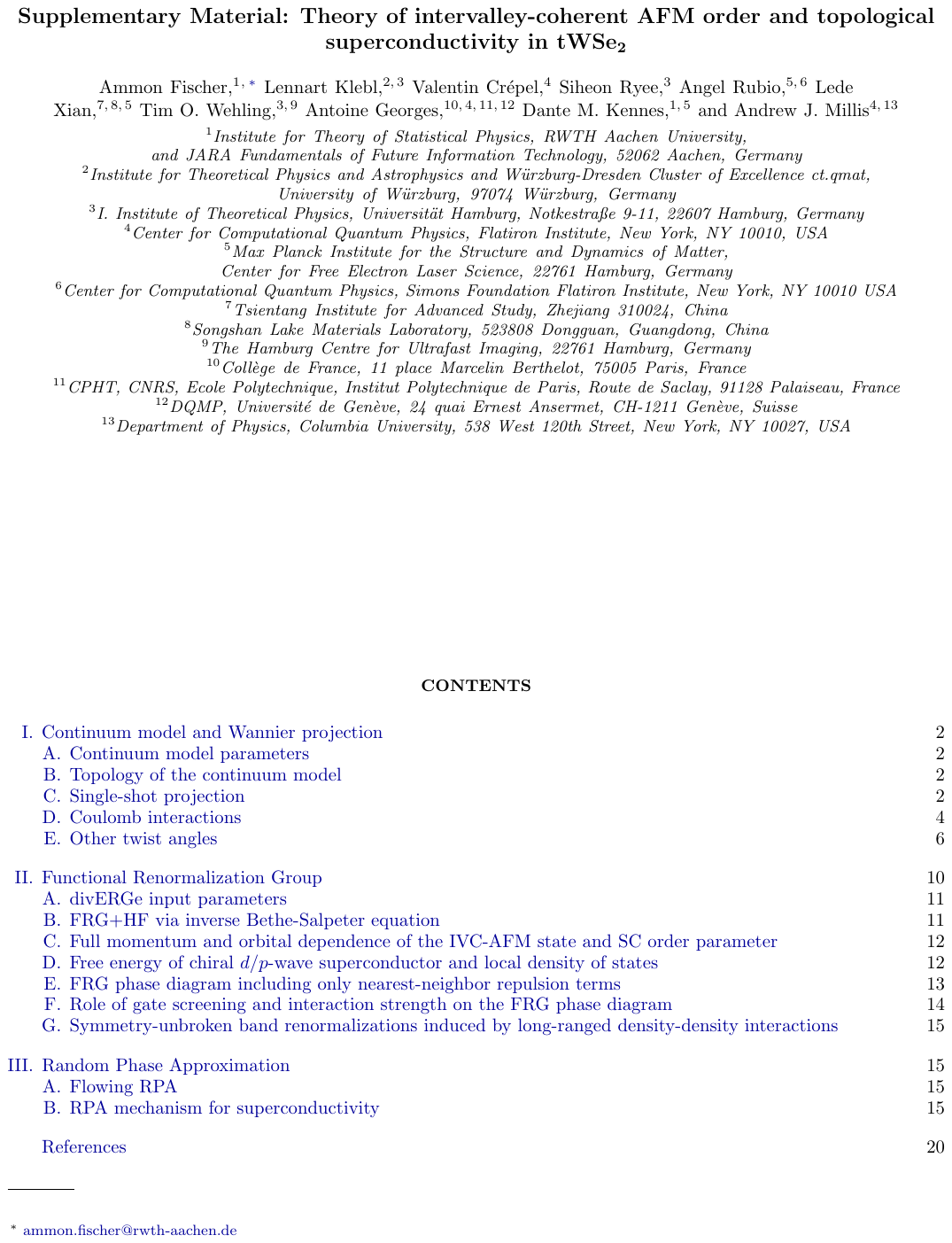} 
}

\end{document}